# Interference Alignment for the MIMO Interference Channel with Delayed Local CSIT

Akbar Ghasemi†, Abolfazl Seyed Motahari‡ and Amir Keyvan Khandani†

†Department of Electrical and & Computer Engineering

University of Waterloo, Waterloo, ON, Canada N2L 3G1

Emails: {aghasemi, khandani}@cst.uwaterloo.ca

‡Department of Electrical Engineering and Computer Sciences

University of California-Berkeley, Berkeley, CA, USA

Email: abolfazl@cst.uwaterloo.ca

**Abstract**

We consider the MIMO (multiple-input multiple-output) Gaussian interference channel with i.i.d. fading across antennas and channel uses and with the delayed local channel state information at the transmitters (CSIT). For the two-user case, achievability results for the degrees of freedom (DoF) region of this channel are provided. We also prove the tightness of our achievable DoF region for some antenna configurations. Interestingly, there are some cases in which the DoF region with delayed local CSIT is identical to the DoF region with perfect CSIT and that is strictly larger than the DoF region with no CSIT. We then consider the $K$-user MISO (multiple-input single-output) IC and show that the degrees of freedom of this channel could be greater than one with delayed local CSIT. *

Financial support provided by Nortel and the corresponding matching funds by the Natural Sciences and Engineering Research Council of Canada (NSERC), and Ontario Centers of Excellence (OCE) are gratefully acknowledged.

*At the final stage of this work, we realized that some parts of the material of this work have been independently reported in [11].

# I. INTRODUCTION

Interference alignment (IA) is an effective technique to mitigate the severe effect of interference in multi-user channels where several transmitter-receiver pairs share the same communication medium. In its original form [1], [2], IA requires the perfect and instantaneous channel state information (CSI) at all nodes to reveal its full potential. The availability of perfect and instantaneous CSI at the receivers (CSIR) can be realized in practice by channel estimation. The perfect and instantaneous CSI at the transmitters (CSIT), however, is hard to obtain in practice. To overcome this problem, one needs to consider the possibility of IA with no/partial CSIT. Considering degrees of freedom (DoF) as the performance measure, it has been recently approved that with the independent and identically distributed (i.i.d.) Rayleigh fading for channel coefficients across time and space, the advantage of IA collapses with no-CSIT for some multi-user channels like MIMO broadcast channel (BC) [3], [4] or two-user MIMO interference channel (IC) [3]–[5]. On the other hand, under channel correlation assumption, the possibility of IA with no-CSIT has been demonstrated in [6]. Recently, in [7], Maddah-Ali and Tse introduced a new model for the availability of CSI in the context of MISO BC which is interesting from both theoretical and practical standpoints. In this model which is commonly referred to as *delayed CSIT*, channel coefficients experience i.i.d. Rayleigh fading across antennas and channel uses. Moreover, each receiver knows its own channel matrices perfectly and instantaneously while all other nodes know it with a unit delay. The remarkable finding of [7] is that the DoF of a MISO BC channel with delayed CSIT can be strictly greater than that with no CSIT. In [8], Maleki *et. al.* have extended the idea of [7] to the more distributed cases like ICs and $X$ channels. Very recently, Vaze and Varanasi have characterized the DoF region of the two-user MIMO BC with delayed CSIT in [9] which is a generalization of the result in [7] for the two-user case. In this paper, we consider the MIMO Gaussian interference channel under *delayed local CSIT* assumption. Similar to the delayed CSIT model, in the delayed local CSIT model, channel coefficients experience i.i.d. Rayleigh fading across antennas and channel uses. Moreover, each receiver knows its own channel matrices perfectly and instantaneously while all other *receivers* know it perfectly but with a unit delay. Unlike the delayed CSIT model in which each transmitter knows the global CSI perfectly and with a unit delay, in the delayed local CSIT model, each transmitter knows its *own channel matrices* perfectly and with a unit delay. We first consider the two-user MIMO Gaussian IC under delayed local CSIT assumption. We provide achievability results on the DoF region of this channel. We then show that our achievable scheme is tight for some antenna configurations. Similar to the result of [7], our results indicate the advantage of delayed local CSIT compared with the no CSIT situation for the two-user MIMO IC. Next, we consider the $K$-user MISO Gaussian IC with $M$ antennas at each transmitter and under delayed local CSIT assumption. We show that when $M \geq K$, we can achieve a sum-DoF which is strictly greater than one. This is in sharp contrast to the no CSIT case where the sum-DoF collapses to one [3]. This shows that even delayed local CSIT can be quite useful in achieving higher sum-DoF for $K > 2$ user MISO Gaussian IC.

The rest of this paper is organized as follows: In section II, the system model is described and the main results are presented. Next, we prove our achievability results for the two-user case in sections III, IV, and V. We then prove that our achievable scheme for the two-user case is tight for some special cases in section IV. We provide achievability as well as upper-bound results on the sum-DoF of the $K$-user MISO Gaussian IC in section VII. We conclude in section VIII.

## II. System Model and Main Results

### A. The Two-User MIMO Gaussian IC

Consider the two-user MIMO Gaussian IC with $M_1$, $M_2$ antennas at the transmitters and $N_1$, $N_2$ antennas at their corresponding receivers. The input-output relationship of this channel can be described as

$$\mathbf{Y}^{[k]}(t) = \mathbf{H}^{[k1]}(t)\mathbf{X}^{[1]}(t) + \mathbf{H}^{[k2]}(t)\mathbf{X}^{[2]}(t) + \mathbf{Z}^{[k]}(t),\ k = 1, 2$$

where at time index $t$, $\mathbf{X}^{[j]}(t) \in \mathbb{C}^{M_j}$ is the transmit signal of user $j$, $\mathbf{Y}^{[k]}(t) \in \mathbb{C}^{N_k}$ is the received signal at receiver $k$, $\mathbf{H}^{[kj]}(t) \in \mathbb{C}^{N_k \times M_j}$ is the channel matrix between transmitter $j$ and receiver $k$ and $\mathbf{Z}^{[k]}(t) \in \mathbb{C}^{N_k}$ is the complex Additive White Gaussian Noise (AWGN) vector at receiver $k$. The transmitters are required to satisfy the same power constraint $\mathbb{E}[||\mathbf{X}^{[k]}||^2] \leq P,\ k = 1, 2$. We further assume that the channel matrices experience independent and identically distributed (i.i.d.) Rayleigh fading across time and space and are independent of receiver noises. That is the elements of $\mathbf{H}^{[kj]}(t)$ are i.i.d. standard complex Gaussian random variables across time and space.

A rate pair $(R_1(P), R_2(P))$ is said to be achievable for the two-user MIMO Gaussian IC if the transmitters can increase the cardinalities of their message sets as $2^{nR_i(P)}$ with block length $n$ and the average probability of error for both users can be made arbitrarily small when $n$ is sufficiently large. The capacity region $\mathscr{C}(P)$ of the two-user MIMO Gaussian IC is the set of all achievable rate pairs $(R_1(P), R_2(P))$. Let $\mathbb{R}_+$ denote the set of all non-negative real numbers. The DoF region $\mathscr{D}$ of the two-user MIMO Gaussian IC is the set of all pairs $(d_1, d_2) \in \mathbb{R}_+^2$ for them there exist a rate pair $(R_1(P), R_2(P))$ in $\mathscr{C}(P)$ such that $d_i = \lim_{P\to\infty} \frac{R_i(P)}{\log(P)},\ i = 1, 2$.

It is assumed that each receiver has access to its own channel matrices perfectly and instantaneously while the other receiver knows it perfectly but with a unit delay. Moreover, each transmitter knows its own channel matrices perfectly but with a unit delay. More precisely, at time index $t$, receiver $k$ has access to $\{\mathbf{H}^{[k1]}(t'), \mathbf{H}^{[k2]}(t')\}_{t'=1}^{t}$ and $\{\mathbf{H}^{[\bar{k}1]}(t'), \mathbf{H}^{[\bar{k}2]}(t')\}_{t'=1}^{t-1}$, $\bar{k} = \{1, 2\} \setminus k$, and transmitter $k$ has access to $\{\mathbf{H}^{[1k]}(t'),\mathbf{H}^{[2k]}(t')\}(t')\}_{t'=1}^{t-1}$. This assumption about the CSI knowledge will be referred to as *delayed local CSIT*. In the following, the DoF region of the two-user MIMO IC with delay local CSIT will be denoted by $\mathscr{D}_{\text{IC}}^{\text{d-CSI}}$.

### B. The $K$-user MISO Gaussian IC

Consider the $K$-user MISO Gaussian IC with $M$ antennas at each transmitter. The input-output relationship of this channel can be describe by

$$y^{[k]}(t)=\mathbf{H}^{[k1]}(t)\mathbf{X}^{[1]}(t)+\mathbf{H}^{[k2]}(t)\mathbf{X}^{[2]}(t)+z^{[k]}(t),\ k=1,\cdots,K$$

where at time index $t$, $\mathbf{X}^{[j]}(t)\in\mathbb{C}^M$ is the transmit signal of user $j$, $y^{[k]}(t)\in\mathbb{C}$ is the received signal at receiver $k$, $\mathbf{H}^{[kj]}(t)\in\mathbb{C}^{1\times M}$ is the channel matrix between transmitter $j$ and receiver $k$ and $z^{[k]}(t)\in\mathbb{C}$ is the complex Additive White Gaussian Noise (AWGN) at receiver $k$. All transmitters are required to satisfy the same power constraint $\mathbb{E}[||\mathbf{X}^{[k]}||^2]\leq P$, $k=1,\cdots,K$. We further assume that the channel matrices experience independent and identically distributed (i.i.d.) Rayleigh fading across time and space and are independent of receiver noises. It is assumed that at time index $t$, receiver $i$ has access to $\{\mathbf{H}^{[i1]}(t'),\mathbf{H}^{[i2]}(t'),\cdots,\mathbf{H}^{[iK]}(t')\}_{t'=1}^{t}$ and $\bigcup_{\bar{i}\neq i}\{\mathbf{H}^{[\bar{i}1]}(t'),\mathbf{H}^{[\bar{i}2]},\cdots,\mathbf{H}^{[\bar{i}K]}(t')\}_{t'=1}^{t-1}$, and transmitter $j$ has access to $\{\mathbf{H}^{[1j]}(t'),\mathbf{H}^{[2j]}(t'),\cdots,\mathbf{H}^{[Kj]}(t')\}_{t'=1}^{t-1}$. The notions of achievable rates, capacity region $\mathscr{C}$ and DoF-region $\mathscr{D}$ can be defined similar to those for the two-user case. The sum-DoF of this channel is defined as $\max_{\mathscr{D}}(d_1+d_2+\cdots+d_K)$.

### C. Main Results

Consider the two-user MIMO Gaussian IC with $M_1,M_2$ antennas at the transmitters and $N_1,N_2$ antennas at their corresponding receivers. Without loss of generality we will assume $N_2\geq N_1$. The results for the case of $N_2<N_1$ can be easily obtained by exchanging the users' indices. Assuming $N_2\geq N_1$, we consider the following six possibilities for different values of $M_1,N_1,M_2$, and $N_2$. One can easily see that these six classes include every antenna configuration and moreover they are mutually exclusive:

$$\begin{aligned}
\mathcal{C}_1 &= \{(M_1, M_2, N_1, N_2) | M_2 \le N_1\} \\
\mathcal{C}_2 &= \{(M_1, M_2, N_1, N_2) | M_2 > N_1 \text{ and } M_1 > N_1 \text{ and } M_2 \le N_2\} \\
&= \{(M_1, M_2, N_1, N_2) | \min(M_1, M_2) > N_1 \text{ and } M_2 \le N_2\} \\
\mathcal{C}_3 &= \{(M_1, M_2, N_1, N_2) | M_2 > N_1 \text{ and } M_1 > N_1 \text{ and } M_2 > N_2 \text{ and } M_1 \le N_2\} \\
&= \{(M_1, M_2, N_1, N_2) | N_1 < M_1 \le N_2 < M_2\} \\
\mathcal{C}_4 &= \{(M_1, M_2, N_1, N_2) | M_2 > N_1 \text{ and } M_1 > N_1 \text{ and } M_2 > N_2 \text{ and } M_1 > N_2\} \\
&= \{(M_1, M_2, N_1, N_2) | \min(M_1, M_2) > N_2 \ge N_1\} \\
\mathcal{C}_5 &= \{(M_1, M_2, N_1, N_2) | M_2 > N_1 \text{ and } M_1 \le N_1 \text{ and } M_2 \le N_2\} \\
&= \{(M_1, M_2, N_1, N_2) | M_1 \le N_1 < M_2 \le N_2\} \\
\mathcal{C}_6 &= \{(M_1, M_2, N_1, N_2) | M_2 > N_1 \text{ and } M_1 \le N_1 \text{ and } M_2 > N_2\} \\
&= \{(M_1, M_2, N_1, N_2) | M_1 \le N_1 \le N_2 < M_2\}
\end{aligned} \tag{1}$$

First, we need the following definition.

*Definition 1:* The region $\mathscr{D}_{\text{IC,in}}^{\text{d-CSI}}$ is defined as

$$\mathscr{D}_{\text{IC,in}}^{\text{d-CSI}} = \left\{ (d_1, d_2) \in \mathbb{R}_+^2 \;\middle|\; \begin{array}{c} d_1 \le M_1 \\ d_2 \le M_2 \\ \frac{d_1}{\max(M_1', N_2)} + \frac{d_2}{N_2} \le 1 \\ \frac{d_1}{N_1} + \frac{d_2}{\max(M_2', N_1)} \le 1 \end{array} \right\}, \tag{2}$$

where $M_i' = \min(M_i, N_1 + N_2)$, $i = 1, 2$.

The following theorem provides an inner-bound for the DoF region of the two-user MIMO IC with delayed local CSI for all antenna configurations except a subclass of class $\mathcal{C}_6$.

*Theorem 1:* The DoF region of the two-user MIMO Gaussian IC with delayed local CSIT and with $N_2 \ge N_1$ contains $\mathscr{D}_{\text{IC,in}}^{\text{d-CSI}}$ for all antenna configurations except for a subclass $\mathcal{S}$ of class $\mathcal{C}_6$ defined by

$$\mathcal{S} \triangleq \{(M_1, M_2, N_1, N_2) | \Delta < M_1 < N_1 < N_2 < N_1 + N_2 - M_1 < M_2\}, \tag{3}$$

where $\Delta = \frac{N_1(N_1 - M_1)}{N_2 - M_1}$.

*Proof:* See section III. ■

We should point out here that some of the inequalities in (2) may be inactive for some antenna configurations.

The subclass $\mathcal{S}$ described in Theorem 1, can be further subdivided into two disjoint subclasses $\mathcal{S}_1$ and $\mathcal{S}_2$ defined as

$$\begin{aligned} \mathcal{S}_1 &\triangleq \{(M_1, M_2, N_1, N_2) \in \mathcal{S} |\, M_2 \leq N_1 + N_2 - \Delta\} \\ \mathcal{S}_2 &\triangleq \{(M_1, M_2, N_1, N_2) \in \mathcal{S} |\, M_2 > N_1 + N_2 - \Delta\}\,. \end{aligned} \tag{4}$$

We have the following inner-bounds for DoF region in subclass $\mathcal{S}_1$ and $\mathcal{S}_2$.

*Theorem 2:* For subclass $\mathcal{S}_1$ of the two-user MIMO Gaussian ICs with delayed local CSIT, the DoF region contains $\mathscr{D}^{\text{d-CSI}}_{\text{IC,in}} \bigcap \mathcal{L}$ where region $\mathcal{L}$ is described by

$$\mathcal{L} = \{(d_1, d_2) \in \mathbb{R}^2_+ |\, \frac{d_1}{\alpha} + \frac{d_2}{N_1 + N_2} \leq 1\}, \tag{5}$$

and $\alpha = \frac{N_1(N_1+N_2)}{2N_1+N_2-M_1}$.

*Proof:* See section IV. ■

*Theorem 3:* For subclass $\mathcal{S}_2$ of the two-user MIMO Gaussian ICs with delayed local CSIT, the DoF region contains the pentagon with corner points $(0,0)$, $(M_1, 0)$, $(0, N_2)$, $(\frac{N_1^2}{M_2-\lambda}, N_2 - \frac{N_1^2}{M_2-\lambda})$, and $(M_1, \frac{(M_2-\lambda)(N_1-M_1)}{N_1})$ where $\lambda = M_1 + M_2 - N_1 - N_2$.

*Proof:* See section V. ■

*Theorem 4:* The achievable DoF regions described in Theorem 1-3 are tight in the following cases:

a) $M_2 \leq N_1$
b) $N_1 < M_1 \leq N_2$ and $M_2 \geq N_1 + N_2$
c) $\min(M_1, M_2) \geq N_1 + N_2$
d) $M_1 \leq \Delta < N_1 \leq N_2 < N_1 + N_2 - M_1 < M_2$
e) $M_1 \leq \Delta' < N_1 \leq N_2 < M_2 \leq N_1 + N_2 - M_1$,

where $\Delta' = \frac{N_1(M_2-N_2)}{M_2-N_1}$.

*Proof:* See section VI. ■

*Theorem 5:* The DoF of $K$-user MISO IC with $M \geq K$ antennas at each transmitter and with delayed local CSIT is greater than or equal to $\frac{K}{K^2-K+1}K$ and is less than or equal to $\frac{K^2-2K+2}{K^2-K+1}K$.

*Proof:* See section VII. ■

### D. Discussions

To examine our achievable DoF region for the two-use MIMO IC in Theorem 1-3, we consider each of the six classes defined in (1) separately. For each class, we compare our achievable DoF region under delayed local CSIT assumption with DoF region under perfect CSIT and no CSIT assumption.

- **Class** $\mathcal{C}_1$: $M_2 \leq N_1$

  For this class, the third inequality in (2) is inactive. It has been proved in [3], [4] that for this class

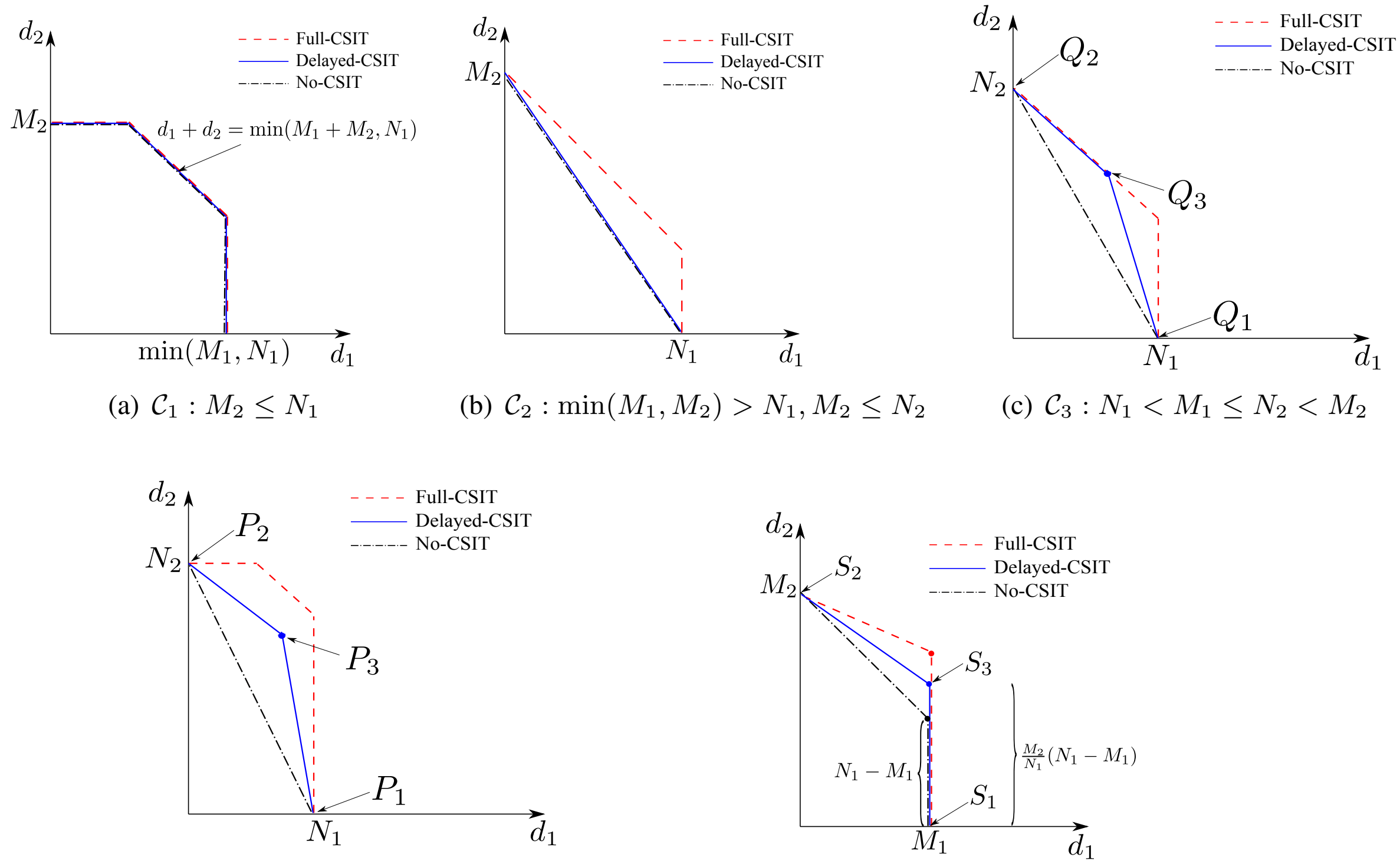

(a) $\mathcal{C}_1 : M_2 \leq N_1$ (b) $\mathcal{C}_2 : \min(M_1, M_2) > N_1, M_2 \leq N_2$ (c) $\mathcal{C}_3 : N_1 < M_1 \leq N_2 < M_2$

(d) $\mathcal{C}_4 : \min(M_1, M_2) > N_2 \geq N_1$ (e) $\mathcal{C}_5 : M_1 \leq N_1 < M_2 \leq N_2$

Fig. 1. The achievable DoF region for the two-user MIMO Gaussian IC with $N_2 \geq N_1$ and with delayed local CSIT. The DoF region of the same channel with no CSIT and full CSIT are also presented for comparison.

the DoF region with no CSIT coincides with DoF region with perfect CSIT and that is described by (2). Therefore, the region $\mathscr{D}_{\text{in}}^{\text{d-CSI}}$ in (2) is also the DoF region for the delayed local CSIT case. A typical shape of DoF region for this class is depicted in Fig. 1(a).

- **Class** $\mathcal{C}_2$: $\min(M_1, M_2) > N_1$ and $M_2 \leq N_2$

  For this class, the first three inequalities in (2) are inactive and the region $\mathscr{D}_{\text{in}}^{\text{d-CSI}}$ in (2) is reduced to the time division region $\frac{d_1}{N_1} + \frac{d_2}{M_2} \leq 1$. It has been shown in [3], [4] that the time division region is indeed equal to the DoF region with no CSIT and is strictly smaller than the DoF region with perfect CSIT. This is the only class that our achievable DoF region with the delayed local CSIT is strictly smaller than the DoF region with perfect CSIT and is not larger than the the DoF region with no CSIT. A typical shape of DoF region for this class is depicted in Fig. 1(b).

- **Class** $\mathcal{C}_3 : N_1 < M_1 \leq N_2 < M_2$

  For this class, the first two inequalities in (2) are inactive. One can check that under the conditions of this class we have $\max(M_1', N_2) = N_2$ and $\max(M_2', N_1) = M_2'$. Our achievable DoF region is then a quadrilateral with corner points $Q_0 = (0, 0)$, $Q_1 = (N_1, 0)$, $Q_2 = (0, N_2)$, and $Q_3 = (\frac{N_1(M_2'-N_2)}{M_2'-N_1}, \frac{M_2'(N_2-N_1)}{M_2'-N_1})$. The DoF region with no CSIT is equal to the time-division region for this class [3], [4]. A typical shape of DoF region for this class is depicted in Fig. 1(c).

- **Class** $\mathcal{C}_4 : \min(M_1, M_2) > N_2 \geq N_1$

  For this class, the first two inequalities in (2) are inactive. It is straightforward to check that under the conditions of this class we have $\max(M_1', N_2) = M_1'$ and $\max(M_2', N_1) = M_2'$. Therefore, our achievable DoF region is a quadrilateral with corner points $P_0 = (0, 0)$, $P_1 = (N_1, 0)$, $P_2 = (0, N_2)$, and $P_3 = (\frac{M_1' N_1 (M_2' - N_2)}{M_1' M_2' - N_1 N_2}, \frac{M_2' N_2 (M_1' - N_1)}{M_1' M_2' - N_1 N_2})$. The time-division region is again equal to the DoF region with no CSIT [3], [4]. A typical shape of DoF region for this class is depicted in Fig. 1(d).
- **Class** $\mathcal{C}_5 : M_1 \leq N_1 < M_2 \leq N_2$

  For this class, $\max(M_1', N_2) = N_2$ and $\max(M_2', N_1) = M_2$ and hence the second and the third inequalities in (2) are inactive. Our achievable DoF region is a trapezoid with corner points $S_0 = (0, 0)$, $S_1 = (M_1, 0)$, $S_2 = (0, M_2)$, and $S_3 = (M_1, \frac{M_2(N_1 - M_1)}{N_1})$. It has been proved in [5] that for this class the DoF region with no CSIT is also a trapezoid with corner points $(0, 0)$, $(M_1, 0)$, $(M_1, N_1 - M_1)$, and $(0, M_2)$. A typical shape of DoF region for this class is depicted in Fig. 1(e).
- **Class** $\mathcal{C}_6 : M_1 \leq N_1 \leq N_2 < M_2$

  It has been proved in [5] that for this class the DoF region with no CSIT is a trapezoid with corner points $T_0 = (0, 0)$, $T_1 = (M_1, 0)$, $T_2 = (0, N_2)$, $T_3 = (M_1, N_1 - M_1)$. With perfect CSIT, the DoF region is also a trapezoid with corner points $T_0$, $T_1$, $T_2$, and $T_4 = (M_1, N_2 - M_1)$. To examine our DoF region for the delayed local CSIT, we need to divide this class to the disjoint union of four subclasses:

$$\mathcal{C}_6 = \mathcal{C}_{61} \bigcup \mathcal{C}_{62} \bigcup \mathcal{C}_{63} \bigcup \mathcal{S},$$

  where $\mathcal{S}$ was defined in (3) and $\mathcal{C}_{61}, \mathcal{C}_{62}$, and $\mathcal{C}_{63}$ are defined as

$$\mathcal{C}_{61} = \{(M_1, M_2, N_1, N_2) | M_1 \leq \Delta < N_1 \leq N_2 < N_1 + N_2 - M_1 < M_2\}$$

$$\mathcal{C}_{62} = \{(M_1, M_2, N_1, N_2) | M_1 \leq \Delta' < N_1 < N_2 < M_2 \leq N_1 + N_2 - M_1\}$$

$$\mathcal{C}_{63} = \{(M_1, M_2, N_1, N_2) | \Delta' < M_1 < N_1 < N_2 < M_2 \leq N_1 + N_2 - M_1\}$$

  Subclass $\mathcal{S}$ can be further subdivided into two subclasses $\mathcal{S}_1$ and $\mathcal{S}_2$ as defined in (4). For subclass $\mathcal{C}_{61}$ and subclass $\mathcal{C}_{62}$, the second and fourth inequalities in (2) are inactive and our achievable DoF region is a trapezoid with corner points $T_0$, $T_1$, $T_2$, and $T_4$ which is equal to the DoF region with perfect CSIT. Therefore our achievable DoF region is tight for subclass $\mathcal{C}_{61}$ and subclass $\mathcal{C}_{62}$. A typical shape of DoF region for these subclasses is depicted in Fig. 2(a) . For subclass $\mathcal{C}_{63}$, the second inequality in (2) is inactive and our achievable DoF region is a pentagon with corner points $T_0$, $T_1$, $T_2$, $T_5 = (\frac{N_1(M_2 - N_2)}{M_2 - N_1}, \frac{M_2(N_2 - N_1)}{M_2 - N_1})$, and $T_6 = (M_1, \frac{M_2(N_1 - M_1)}{N_1})$. A typical shape of DoF region for this case is depicted in Fig. 2(b) . From Theorem 2, our achievable DoF region for subclass $\mathcal{S}_1$ is a hexagon

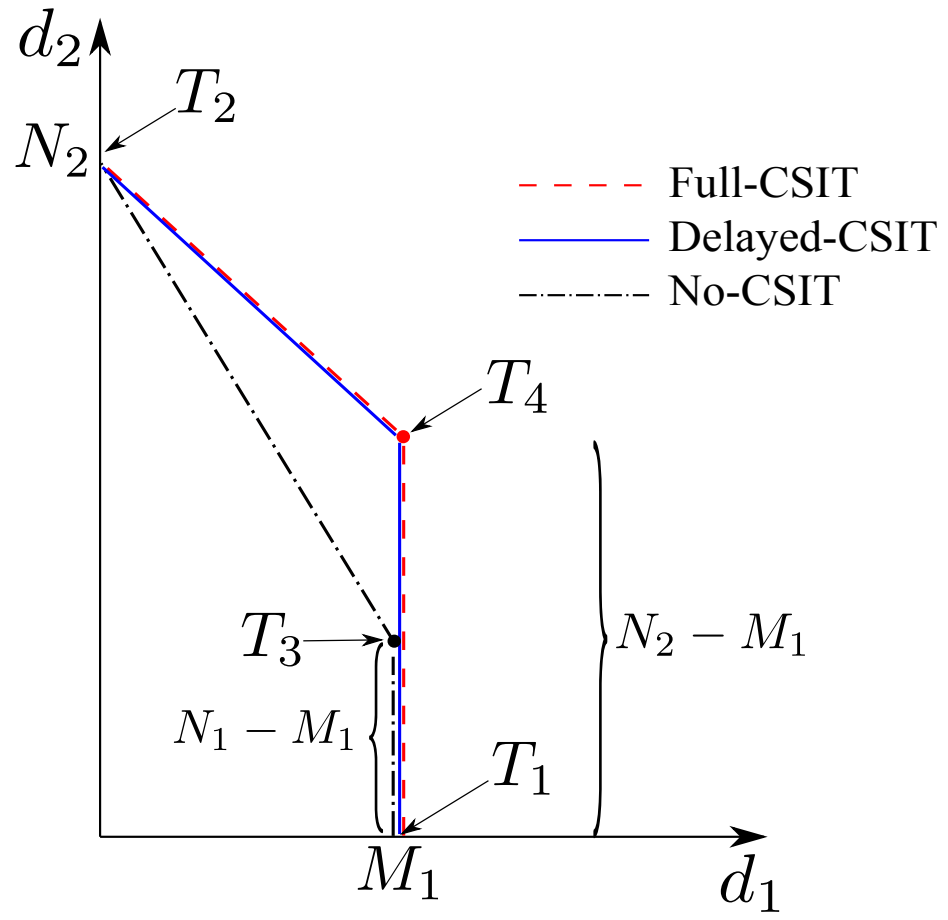

(a) Subclass $\mathcal{C}_{61}$ and $\mathcal{C}_{62}$

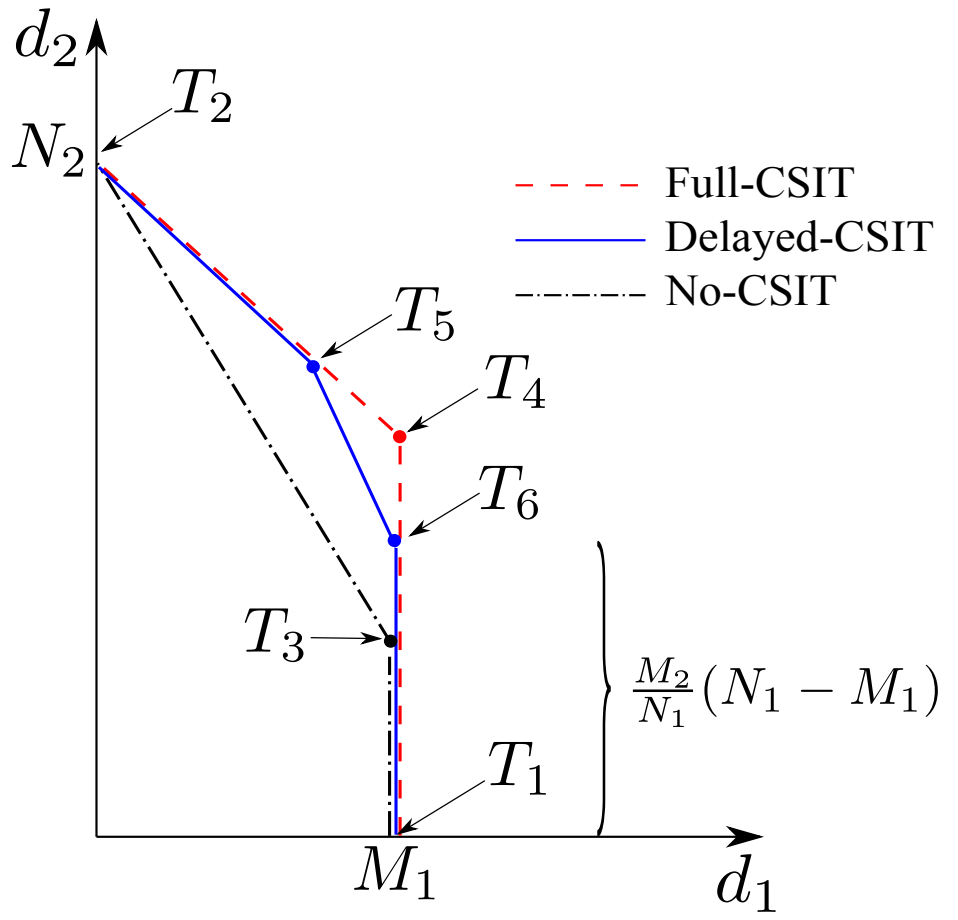

(b) Subclass $\mathcal{C}_{63}$

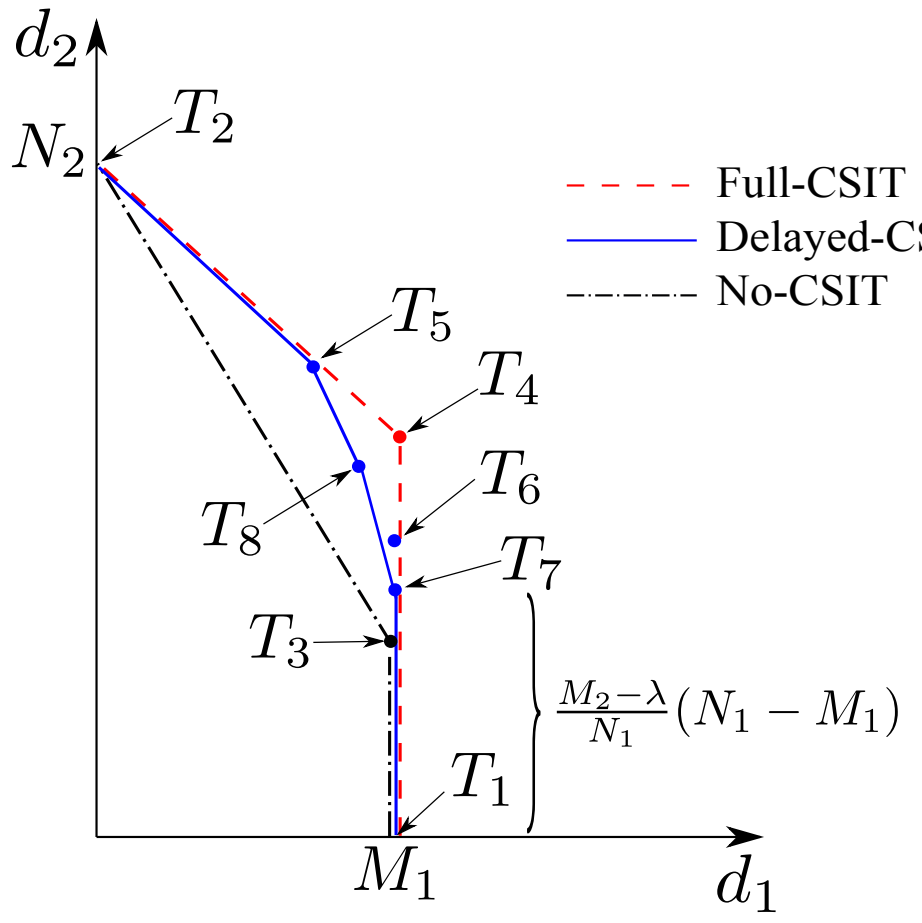

(c) Subclass $\mathcal{S}_1$

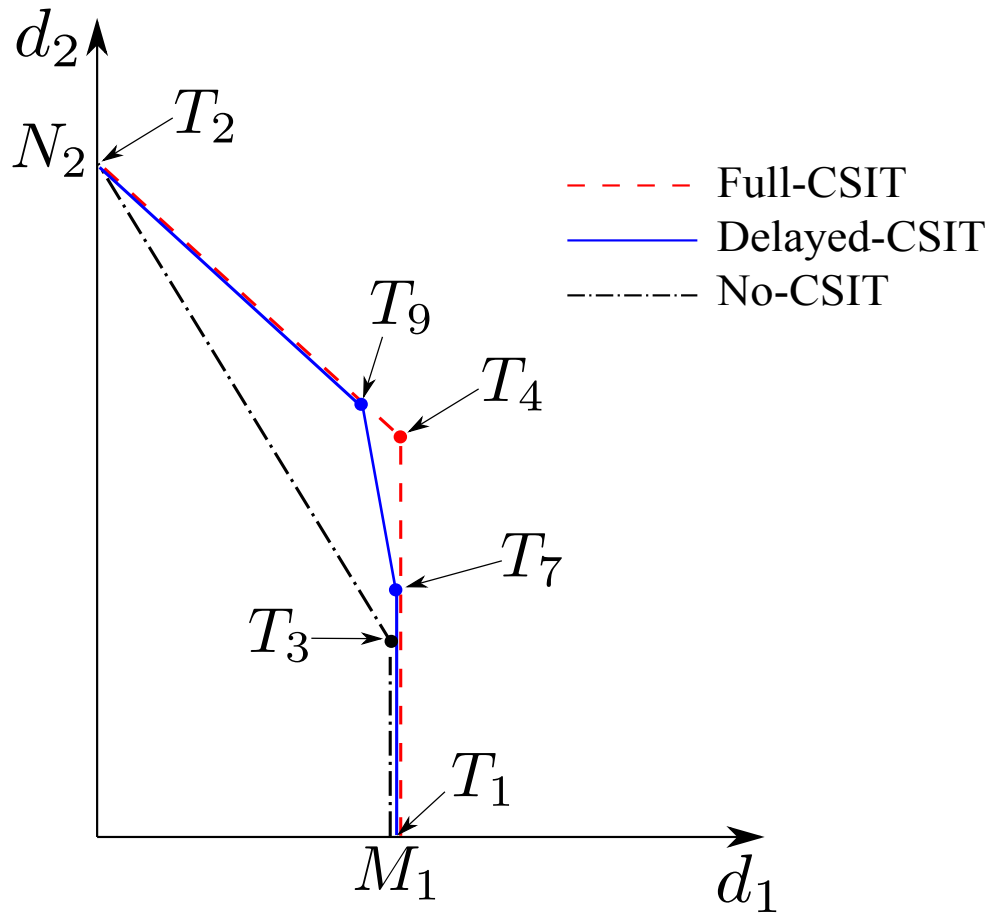

(d) Subclass $\mathcal{S}_2$

Fig. 2. The achievable DoF region for the two-user MIMO Gaussian IC with $N_2 \geq N_1$ and with delayed local CSIT for class $\mathcal{C}_6$ ($M_1 \leq N_1 \leq N_2 < M_2$). The DoF region of the same channel with no CSIT and full CSIT are also presented for comparison.

with corner points $T_0$, $T_1$, $T_2$, $T_5$, $T_7 = (M_1, \frac{(M_2-\lambda)(N_1-M_1)}{N_1})$, and $T_8 = (\frac{N_1(M_1-\lambda)}{N_1-\lambda}, \frac{M_2(N_1-M_1)}{N_1-\lambda})$. A typical shape of our achievable DoF region for subclass $\mathcal{S}_1$ is depicted in Fig. 2(c). Finally, as stated in Theorem 3, our achievable DoF region for subclass $\mathcal{S}_2$ is a pentagon with corner points $T_0$, $T_1$, $T_2$, $T_7$, and $T_9 = (\frac{N_1^2}{M_2-\lambda}, N_2 - \frac{N_1^2}{M_2-\lambda})$. A typical shape of our achievable DoF region for subclass $\mathcal{S}_2$ is shown in Fig. 2(d).

Theorem 2 identifies some configurations for them our achievable DoF region gives an exact characterization of DoF region. This includes class $\mathcal{C}_1$, a subclass of class $\mathcal{C}_3$, a subclass of class $\mathcal{C}_4$, and subclasses $\mathcal{C}_{61}$ and $\mathcal{C}_{62}$ of class $\mathcal{C}_6$.

Here, an interesting observation is that for all antenna configurations except class $\mathcal{C}_4$, the sum-DoF of the two-user MIMO IC does not change with the knowledge of CSIT. For class $\mathcal{C}_4$, however, the sum-DoF of the two-user MIMO IC with full CSIT could be strictly greater than that with delayed local CSIT which in turn is strictly greater than the sum-DoF with no CSIT. Specifically, when $\min(M_1, M_2) \geq N_1 + N_2$, the sum-DoF with no CSIT is equal to $\max(N_1, N_2)$ while with delayed local CSIT it is equal to $(1 - \frac{N_1 N_2}{N_1^2 + N_2^2 + N_1 N_2})(N_1 + N_2)$, and with full CSIT it is equal to $N_1 + N_2$. Using these results, we can develop a new upper-bound on the DoF of the $K$-user MISO interference channel with $M \geq K$ antennas at each transmitter and with delayed local CSIT as stated in theorem 5. By extending our achievability scheme to the $K$-user MISO IC with $M \geq K$, we can achieve a sum-DoF which is strictly greater than one. It is important to notice that the sum-DoF of this channel with no CSIT is equal to one [3]. This shows that even delayed local CSIT can be quite useful in achieving higher DoF for the $K$-user MISO IC.

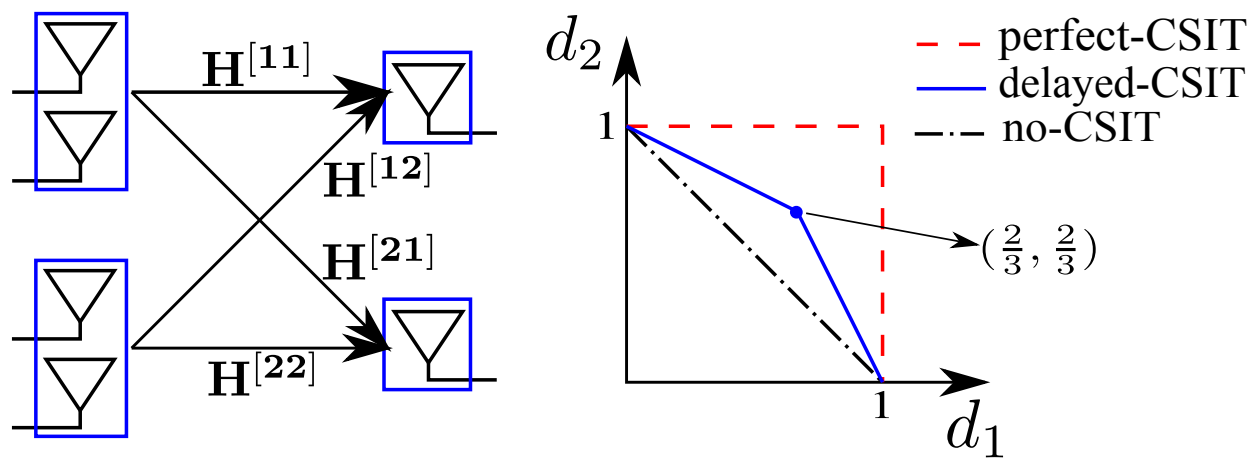

Fig. 3. The DoF region of the two-user MIMO IC with $M_1 = M_2 = 2$ and $N_1 = N_2 = 1$ and under different assumption on CSIT information.

## III. Proof of Theorem 1

In this section, we prove that the region stated in Theorem 1 is achievable. For illustration purpose, we first elaborate our achievable scheme for a two-user MIMO IC with two antennas at each transmitter and a single antenna at each receiver. We then prove our achievable scheme for general setting.

### A. An illustrative example

Consider a two-user MIMO IC with $M_1 = M_2 = 2$ and $N_1 = N_2 = 1$. This channel is depicted in Fig. 3. We first notice that the DoF region with perfect CSIT is the unit square $d_i \leq 1$, $i = 1, 2$. Also, the DoF region with no CSIT is the time division region described by $d_1 \leq 1, d_2 \leq 1$, $d_1 + d_2 \leq 1$. These regions are depicted in Fig. 3. Now, we show that the DoF region with delayed local CSIT is strictly larger than the DoF region with no CSIT and is strictly smaller than the DoF region with perfect CSIT.

We first show that the point $(d_1, d_2) = (2/3, 2/3)$ is achievable with delayed local CSIT. To this end, we consider the channel over three channel uses. We divide the duration of three channel uses into two phases:

<u>Phase One:</u> This phase takes one channel use. At this phase, each transmitter sends two independent coded symbols for its intended receiver. Specifically, let us assume that transmitter one sends the symbol $u_i^{[1]}, i = 1, 2$ from its $i^{\text{th}}$ transmit antenna while transmitter two sends $u_j^{[2]}, j = 1, 2$ from its $j^{\text{th}}$ transmit antenna. Since we are primarily concerned with the DoF, we can safely disregard the thermal noise at the receiver side [7]. The following signals are appeared at the receivers at the end of first channel use:

$$\begin{aligned} y^{[1]}(1) &= h_{11}^{[11]}(1)u_1^{[1]} + h_{12}^{[11]}(1)u_2^{[1]} + h_{11}^{[12]}(1)u_1^{[2]} + h_{12}^{[12]}(1)u_2^{[2]} \\ y^{[2]}(1) &= h_{11}^{[21]}(1)u_1^{[1]} + h_{12}^{[21]}(1)u_2^{[1]} + h_{11}^{[22]}(1)u_1^{[2]} + h_{12}^{[22]}(1)u_2^{[2]} \end{aligned}$$

Let us define the following notations:

$$\begin{aligned} I^{[1]}(1) &= h_{11}^{[12]}(1)u_1^{[2]} + h_{12}^{[12]}(1)u_2^{[2]} \\ I^{[2]}(1) &= h_{11}^{[21]}(1)u_1^{[1]} + h_{12}^{[21]}(1)u_2^{[1]}. \end{aligned}$$

Since each transmitter has access to the channel coefficients by a unit delay, at the consequent channel uses, transmitter $k$, $k = 1, 2$ has access to $I^{[\bar{k}]}(1)$ where $\bar{k} = \{1, 2\} \setminus k$.

Phase Two: In this phase, we need to deliver both $I^{[1]}(1)$ and $I^{[2]}(1)$ to each receiver. This can be simply performed in two channel uses by time division.

By the end of phase two, receiver one has access to $y^{[1]}(1), I^{[1]}(1)$, and $I^{[2]}(1)$ from which it can extract $u_1^{[1]}$ and $u_2^{[1]}$. Similarly, receiver two can obtain $u_1^{[2]}$ and $u_2^{[2]}$ from $y^{[2]}(1), I^{[1]}(1)$, and $I^{[2]}(1)$. Since each transmitter has sent two independent symbols for its intended receiver in three channel uses, the DoF pair $(2/3, 2/3)$ has been achieved. By time sharing between this point and the trivially achievable points $(1, 0)$ and $(0, 1)$, we reach to the achievable DoF region depicted in Fig. 3. To prove that this region is indeed the DoF region, we allow the transmitters to cooperate. Since cooperation does not reduce capacity, the DoF region of the resulting BC is an outer-bound for the DoF region of the original IC. But the DoF region of BC channel with delayed CSIT has been recently characterized in [9] and for a BC channel with four antennas at the transmitter and a single antenna at each receiver is given by $d_1/2 + d_2 \leq 1$ and $d_1 + d_2/2 \leq 1$. But, this region is exactly equal to our achievable DoF region. This proves the DoF-optimality of our achievable scheme for this special case.

### *B. Proof of Achievability for General setting*

In this part, we prove our achievability result for Class $\mathcal{C}_2$, $\mathcal{C}_3$, $\mathcal{C}_4$, $\mathcal{C}_5$, and subclasses $\mathcal{C}_{61}$ and $\mathcal{C}_{62}$ of class $\mathcal{C}_6$ . We consider $W$ consecutive channel uses. Each transmitter, divides the duration of $W$ channel uses into two phases:

Phase One: For transmitter one, this phase takes $W_1$ channel uses ($W_1 < W$) while for transmitter two it takes $W_2$ channel uses ($W_2 < W$). At each channel use of this phase, transmitter $i$, $i = 1, 2$, sends $M_i'$ independent coded symbols for receiver $i$ where $M_i' = \min(M_i, N_1 + N_2)$. Therefore, by the end of this phase, transmitter $i$, $i = 1, 2$, has sent $W_i M_i'$ independent symbols for its intended receiver. Due to the interference, it is not generally possible for each receiver to resolve its intended symbols. Therefore, we require the second phase.

Phase Two: For transmitter one, this phase takes $W - W_1$ channel uses while for transmitter two it takes $W - W_2$ channel uses. No new information is sent during this phase. At the beginning of this phase, each transmitter is aware of all the interference terms observed by its non-intended receiver during phase one. Therefore, in phase two, each transmitter tries to help its non-intended receiver by sending the linear combinations of these interference terms in each channel use.

Before we proceed further, we need to introduce a few notations. Let $I_n^{[1]}(j)$ denote the interference observed by the $n^{\text{th}}$ receive antenna of user one at $j^{\text{th}}$ channel use of phase one $(1 \leq n \leq N_1, 1 \leq j \leq W_2)$. Similarly, $I_n^{[2]}(j)$ denotes the interference observed by the $n^{\text{th}}$ receive antenna of user two at $j^{\text{th}}$ channel use of phase one $(1 \leq n \leq N_2, 1 \leq j \leq W_1)$. Each term $I_n^{[1]}(j)$ contains $M_2'$ independent variables. Similarly, $I_n^{[2]}(j)$ contains $M_1'$ independent variables. At the $j^{\text{th}}$ channel use of phase one, the first receiver observes $N_1$ interference terms $I_1^{[1]}(j), I_2^{[1]}(j), \cdots, I_{N_1}^{[1]}(j)$ in its receive antennas. All these interference terms have been generated by $M_2'$ independent variables. If $M_2' > N_1$, interference alignment is possible in receiver one and we only need to decode the $N_1$ interference terms in the first receiver. On the other hand, if $M_2' < N_1$, there is no possibility for interference alignment in receiver one and we have to decode all $M_2'$ interference variables. Hence, in general, in each channel use of phase one transmitter two generates $\min(M_2', N_1)$ independent interference quantities in the first receiver. Similarly, transmitter one generates $\min(M_1', N_2)$ independent interference quantities in receiver two in each channel use of its first phase. We assume that no new interference term is generated in the second phase. This can be happened if in phase two each transmitter sends only different linear combinations of the previously observed interference terms in its non-intended receiver during the phase one. That is for example transmitter one sends different linear combinations of $I_n^{[2]}(j)$, $1 \leq j \leq W_1$, $1 \leq n \leq N_2$ in its second phase. In our achievable scheme, each receiver will decode all its intended information symbols and all the interference terms generated by the other transmitter. Under this assumption, by the end of phase two, there are $M_1'W_1$ desired symbols and $\min(M_2', N_1)W_2$ interference terms in the first receiver. Since there are $N_1$ antennas at receiver 1, we have $N_1 W$ dimensions for $W$ channel uses. Therefore, in order to resolve both the desired variables and interference terms in the first receiver, we must have

$$M_1'W_1 + \min(M_2', N_1)W_2 \leq N_1 W. \tag{6}$$

Similarly, a necessary condition to resolve the desired data streams and interference terms in the second receiver is

$$M_2'W_2 + \min(M_1', N_2)W_1 \leq N_2 W. \tag{7}$$

Dividing both sides of (6) and (7) by $W$ and noting that $d_1 = \frac{M_1' W_1}{W}$ and $d_2 = \frac{M_2' W_2}{W}$, we reach to the third and fourth inequalities in (2). We notice that since $\frac{W_i}{W} \leq 1$, $i = 1, 2$, the first and second inequalities in (2) have also been used in our achievability scheme. To complete the proof of our achievability scheme, we need to show that we can find a pair $(W_1, W_2)$ that satisfies (6) and (7) and moreover each receiver can resolve its intended information symbols. To see that this is not always possible with our achievable scheme, we consider the two-user MIMO Gaussian IC with $M_1 = 2, M_2 = 6, N_1 = 3$, and $N_2 = 4$. One can easily check that $W = W_1 = 3$ and $W_2 = 1$ satisfy both (6) and (7). However, as we shall see, we can not achieve $d_1 = \frac{3\times 2}{3}$ and $d_2 = \frac{6\times 1}{3}$ by using our achievable scheme for this case. Our achievable scheme takes three time slots in total. Transmitter one sends two independent symbols in the first channel use ($u_1^{[1]}, u_2^{[1]}$), two independent symbols in the second channel use ($u_3^{[1]}, u_4^{[1]}$), and two independent symbols in the third channel use ($u_5^{[1]}, u_6^{[1]}$). Transmitter two sends six independent symbols in its first channel use ($u_1^{[2]}, \cdots, u_6^{[2]}$) and sends different linear combinations of $I_1^{[1]}(1), I_2^{[1]}(1)$, and $I_3^{[1]}(1)$ in its last two channel uses. One can see that in the last two channel uses, receiver two gets eight equations in the seven variables $u_3^{[1]}, u_4^{[1]}, u_5^{[1]}, u_6^{[1]}, I_1^{[1]}(1), I_2^{[1]}(1)$, and $I_3^{[1]}(1)$. Therefore, receiver two can solve this consistent over-determined system of linear equations to obtain $I_1^{[1]}(1), I_2^{[1]}(1)$, and $I_3^{[1]}(1)$. In its first channel use, receiver two has four equations in eight variables. Obtaining $I_1^{[1]}(1), I_2^{[1]}(1)$, and $I_3^{[1]}(1)$ from phase two, it will have seven equations in eight unknowns which is an under-determined system of linear equations. Therefore, receiver two can not resolve its intended information symbols. In fact, (6) and (7) do not provide sufficient conditions for the achievability of $d_1 = \frac{M_1' W_1}{W}$ and $d_2 = \frac{M_2' W_2}{W}$ in general. As we can see from the above example, the 12 equations in receiver two partitioned into two subsystems of linear equation: an over-determined subsystem and an under-determined subsystem. Such a partitioning of equations can reduce the effective number of equations in the receivers (in the above example the number of effective equations are reduced from 12 to 11). To make sure that each user can resolve the total number of unknown quantities in its intended receiver, the rank of $N_i W$ equations in receiver $i$, $i = 1, 2$ should be equal to the total number of unknown quantities in that receiver. Therefore, we need to calculate the rank of $N_i W$ received equations in receiver $i$, $i = 1, 2$ in our achievable scheme. After $W$ channel uses, receiver $i$ has $N_i W$ linear equations in terms of the elements of the vector $\mathbf{e}_i \triangleq [\mathbf{U}^{[i]}, \mathbf{I}^{[i]}]$ where $\mathbf{U}^{[i]} = [u_1^{[i]}, \cdots, u_{M_i' W_i}^{[i]}]^T$ and $\mathbf{I}^{[i]}$ is a $\min(M_{\bar{i}}', N_i) W_{\bar{i}} \times 1$ column vector containing all the interference quantities that should be resolved at receiver $i$. Let $\mathbf{P}^{[i]}$ denote the coefficient matrix of this set of linear equations for receiver $i$. So, $\mathbf{P}^{[i]}$ is of size $N_i W \times [M_i' W_i + \min(M_{\bar{i}}', N_i) W_{\bar{i}}]$. We Define

$$i_{\max} \triangleq \arg\max_i \{W_i\},$$

$$i_{\min} \triangleq \{1, 2\} \backslash \{i_{\max}\}.$$

In Appendix A, we prove that the rank of $\mathbf{P}^{[i]}$ is given by

$$\text{rank}(\mathbf{P}^{[i]}) = \min\{M'_i W_i + \min(M'_{\bar{i}}, N_i) W_{\bar{i}}, r_1^{[i]} + r_2^{[i]} + r_3^{[i]}\}, \tag{8}$$

where

$$\begin{aligned} r_1^{[i]} &= \min\left\{N_i, M'_i + M'_{\bar{i}}\right\} W_{i_{\min}} \\ r_2^{[i]} &= \min\left\{N_i(W_{i_{\max}} - W_{i_{\min}}), M'_{i_{\max}}(W_{i_{\max}} - W_{i_{\min}}) + \min\left( M'_{i_{\min}}(W_{i_{\max}} - W_{i_{\min}}), M'_{i_{\min}} W_{i_{\min}}, N_{i_{\max}} W_{i_{\min}}\right)\right\} \\ r_3^{[i]} &= \min\left\{N_i(W - W_{i_{\max}}), \min\left( M'_i(W - W_{i_{\max}}), M'_i W_i, N_{\bar{i}} W_i\right) + \min\left( M'_{\bar{i}}(W - W_{i_{\max}}), M'_{\bar{i}} W_{\bar{i}}, N_i W_{\bar{i}}\right)\right\}. \end{aligned}$$

To successfully decode all the required unknowns in both receivers, we need to satisfy the following rank conditions

$$M'_i W_i + \min(M'_{\bar{i}}, N_i) W_{\bar{i}} \leq r_1^{[i]} + r_2^{[i]} + r_3^{[i]}, \; i = 1, 2. \tag{9}$$

To prove our achievability result, we first need to calculate the corner points of region $\mathscr{D}_{\text{IC,in}}^{\text{d-CSI}}$ and find the corresponding $W_1$ and $W_2$. We then need to check whether the resulting $W_1$ and $W_2$ satisfy the rank conditions in (9). In Appendix B, we prove that the corner points of region $\mathscr{D}_{\text{IC,in}}^{\text{d-CSI}}$ satisfy the rank conditions in (9) for class $\mathcal{C}_3$. In Appendix C, D, E, and F we respectively prove a similar result for class $\mathcal{C}_4$, $\mathcal{C}_5$, $\mathcal{C}_{61}$,$\mathcal{C}_{62}$ and $\mathcal{C}_{63}$. This completes the proof of Theorem 1.

## IV. Proof of Theorem 2

In this section, we prove Theorem 2. We first prove that for subclass $\mathcal{S}$ of class $\mathcal{C}_6$, the point $T_7 = (M_1, \frac{(M_2-\lambda)(N_1-M_1)}{N_1})$ is achievable. Remember that for subclass $\mathcal{S}$ of class $\mathcal{C}_6$, we have $\Delta < M_1 < N_1 < N_2 < N_1 + N_2 - M_1 < M_2$. Noting that the value of $\Delta$ is independent of $M_2$, if user two employs only $N_1 + N_2 - M_1$ out of its $M_2$ transmit antennas, the system would be equivalent to a two user MIMO IC which belongs to subclass $\mathcal{C}_{62}$. Therefore, from Theorem 1, one can easily see that point $T_7$ is achievable. To complete the proof of Theorem 2, we need to show that the corner points $T_5 = (\frac{N_1(M_2-N_2)}{M_2-N_1}, \frac{M_2(N_2-N_1)}{M_2-N_1})$ and $T_8 = (\frac{N_1(M_1-\lambda)}{N_1-\lambda}, \frac{M_2(N_1-M_1)}{N_1-\lambda})$ are achievable for subclass $\mathcal{S}_1$. In the following, we explain our achievable scheme for each corner point separately:

Achievability of corner point $T_5$

We consider $W = M_2 - N_1$ consecutive channel uses. Each transmitter has two phases of transmission. The duration of each phase is identical for both transmitters. Phase one takes $W_1 = N_2 - N_1$ time slots and phase two takes $W - W_1 = M_2 - N_2$ time slots. Let us define the following quantities:

$$\begin{aligned} \nu &= \min\{N_1(N_2 - N_1), N_2(M_2 - N_2)\} \\ \nu_1 &= \nu - (N_2 - N_1)(M_2 - N_2) \\ \nu_2 &= N_2(M_2 - N_2) - \nu. \end{aligned} \tag{10}$$

During the phase one, the first transmitter sends random linear combinations of $\nu_1$ independent information symbols from its transmit antennas while transmitter two sends random linear combinations of $M_2(N_2 - N_1)$ independent information symbols from its transmit antennas. One can easily check that under the conditions of subclass $\mathcal{S}_1$ we always have $0 \leq \nu_1 \leq M_1 W_1$. During the Phase two, the first transmitter sends random linear combinations of $\nu_2$ independent information symbols from its transmit antennas while transmitter two sends random linear combinations of the $N_1 W_1$ interference terms observed by receiver one during the first phase. We notice that since $M_2 \leq N_1 + N_2 - \Delta$, we always have $0 \leq \nu_2 \leq M_1(W - W_1)$. Assuming each user can successfully resolve its intended data streams, the first user achieves $d_1 = \frac{\nu_1 + \nu_2}{W} = \frac{N_1(M_2 - N_2)}{M_2 - N_1}$ and the second user achieves $d_2 = \frac{M_2 W_1}{W} = \frac{M_2(N_2 - N_1)}{M_2 - N_1}$ which is the desired result. Therefore, to complete the proof, we need to show that each user can resolve its intended information symbols in the above-described scheme. To this aim, we first required to prove that the total number of unknown in each receiver is less than or equal to the number of equations. At receiver one, we have a total number of $\nu_1 + \nu_2 + N_1 W_1 = N_1(M_2 - N_1)$ unknown quantities and a total number of $N_1 W = N_1(M_2 - N_1)$ equations. At receiver two, we have a total number of $\nu_1 + \nu_2 + M_2 W_1 = N_2(M_2 - N_1)$ unknown quantities and a total number of $N_2 W = N_2(M_2 - N_1)$ equations. Thus, at each receiver the total number of equations is equal to the total number of unknowns. One can check that at receiver one, the received equations are always linearly independent. At receiver two, however, the received equations are not generally linearly independent. In fact, from Phase two, Since $\nu_2$ was selected in a way that no redundant equation created in receiver two, the received equations in receiver two are also linearly independent. This completes the proof of achievability for corner point $T_5$.

Achievability of corner point $T_8$:

We consider $W = N_1 - \lambda$ consecutive channel uses. Remember that $\lambda = M_1 + M_2 - N_1 - N_2$ for subclass $\mathcal{S}_1$. Each transmitter has two phases of transmission. The duration of each phase is identical for both transmitters. Phase one takes $W_1 = N_1 - M_1$ time slots and phase two takes $W - W_1 = M_1 - \lambda$ time slots. Let us define the following quantities:

$$\begin{aligned} \mu_1 &= (N_1 - M_1)(M_1 - \lambda) \\ \mu_2 &= M_1(M_1 - \lambda). \end{aligned} \tag{11}$$

During the phase one, the first transmitter sends random linear combinations of $\mu_1$ independent information symbols from its transmit antennas while transmitter two sends random linear combinations of $M_2(N_1 - M_1)$ independent information symbols from its transmit antennas. During phase two, the first transmitter sends random linear combinations of $\mu_2$ independent information symbols from its transmit antennas while transmitter two sends random linear combinations of the $N_1W_1$ interference terms observed by receiver one during the first phase. Assuming each user can successfully resolve its intended data streams, the first user achieves $d_1 = \frac{\mu_1+\mu_2}{W} = \frac{N_1(M_1-\lambda)}{N_1-\lambda}$ and the second user achieves $d_2 = \frac{M_2W_1}{W} = \frac{M_2(N_1-M_1)}{N_1-\lambda}$ which is the desired result. Therefore, to complete the proof, we need to show that each user can resolve its intended information symbols in the above-described scheme. To this aim, we first required to prove that the total number of unknown in each receiver is less than or equal to the number of equations. At receiver one, we have a total number of $\mu_1 + \mu_2 + N_1W_1 = N_1(N_1 - \lambda)$ unknown quantities and a total number of $N_1W = N_1(N_1 - \lambda)$ equations. At receiver two, we have a total number of $\mu_1 + \mu_2 + M_2W_1 = N_1(M_1 - \lambda) + M_2(N_1 - M_1)$ unknown quantities and a total number of $N_2W = N_2(N_1 - \lambda)$ equations. Since $M_2 \leq N_1 + N_2 - \Delta$, after some algebraic manipulation, one can prove that $N_1(M_1 - \lambda) + M_2(N_1 - M_1) \leq N_2(N_1 - \lambda)$. Thus, at each receiver the total number of equations is greater than or equal to the total number of unknowns. One can check that at receiver one, the received equations are always linearly independent. At the second receiver, situation is different. At phase two, receiver two observes $N_2(M_1 - \lambda)$ equations in $\mu_2 + N_1(N_1 - M_1)$ unknowns. Since $M_2 \leq N_1 + N_2 - \Delta$, it follows that $\mu_2 + N_1(N_1 - M_1) \leq N_2(M_1 - \lambda)$ and therefore in phase two receiver two can resolve $\mu_2$ symbols from transmitter one as well as $N_1(N_1 - M_1)$ independent linear combinations of its own information symbols. In phase one, receiver two observes $N_2(N_1 - M_1)$ equations in $M_2(N_1 - M_1) + \mu_1$ unknowns. Including $N_1(N_1 - M_1)$ independent linear expressions obtained from phase two, receiver two will have $N_2(N_1-M_1)+N_1(N_1-M_1) = (N_2+N_1)(N_1-M_1)$ equations in $M_2(N_1-M_1)+\mu_1 = (N_1-M_1)(N_1+N_2)$ unknowns. Therefore, receiver two will be able to resolve its desired information symbols. This completes the proof of achievability for corner point $T_8$.

## V. Proof of Theorem 3

To prove Theorem 3, we need to prove that all the corner point mentioned in the Theorem are achievable. This boils down to the achievability of $T_7 = (M_1, \frac{(M_2-\lambda)(N_1-M_1)}{N_1})$ and $T_9 = (\frac{N_1^2}{M_2-\lambda}, N_2 - \frac{N_1^2}{M_2-\lambda})$. The proof of achievability for point $T_7$ is similar to that for subclass $\mathcal{S}_1$ and therefore will not be repeated here. To show that point $T_9$ is achievable, we consider $W = M_2 - \lambda = N_1 + N_2 - M_1$ consecutive channel uses. Each transmitter has two phases of transmission. The duration of each phase is identical for both transmitters. Phase one takes $W_1 = N_2 - M_1$ time slots and phase two takes $W - W_1 = N_1$ time slots.

Let us define the following quantities:

$$\begin{aligned} \eta_1 &= N_1(N_1 - M_1) \\ \eta_2 &= N_1 M_1. \end{aligned} \qquad (12)$$

During the phase one, the first transmitter sends random linear combinations of $\eta_1$ independent information symbols from its transmit antennas. In time slot $t$ of phase one, $1 \leq t \leq W_1$, transmitter two sends $\omega_t$ independent information symbols from its transmit antennas where the integers $\omega_t$, $t = 1, \cdots, W_1$ are selected such that $\omega_t \geq N_2$ for every $t$ and moreover $\sum_{t=1}^{W_1} \omega_t = N_2(M_2 - \lambda) - N_1^2$. Since for subclass $\mathcal{S}_2$ we have $N_2 W_1 \leq N_2(M_2 - \lambda) - N_1^2 \leq M_2 W_1$, such a selection is always possible. During phase two, the first transmitter sends random linear combinations of $\eta_2$ independent information symbols from its transmit antennas while transmitter two sends random linear combinations of the $N_1 W_1$ interference terms observed by receiver one during the first phase. Assuming each user can successfully resolve its intended data streams, the first user achieves $d_1 = \frac{\eta_1 + \eta_2}{W} = \frac{N_1^2}{M_2 - \lambda}$ and the second user achieves $d_2 = \frac{\sum_{t=1}^{W_1} \omega_t}{W} = N_2 - \frac{N_1^2}{M_2 - \lambda}$ which is the desired result. Therefore, to complete the proof, we need to show that each user can resolve its intended information symbols in the above-described scheme. To this aim, we first required to prove that the total number of unknowns in each receiver is less than or equal to the number of equations. At receiver one, we have a total number of $\eta_1 + \eta_2 + N_1 W_1 = N_1(N_1 + N_2 - M_1)$ unknown quantities and a total number of $N_1 W = N_1(N_1 + N_2 - M_1)$ equations. At receiver two, we have a total number of $\eta_1 + \eta_2 + N_2(M_2 - \lambda) - N_1^2 = N_2(N_1 + N_2 - M_1)$ unknown quantities and a total number of $N_2 W = N_2(N_1 + N_2 - M_1)$ equations. Thus, at each receiver the total number of equations is equal to the total number of unknowns. One can check that at receiver one, the received equations are always linearly independent. So, we consider the second receiver. At phase two, receiver two observes $N_2 N_1$ equations in $\eta_2 + N_1(N_2 - M_1) = N_1 N_2$ unknowns. Therefore in phase two receiver two can resolve $\eta_2$ symbols from transmitter one as well as $N_1(N_2 - M_1)$ independent linear combinations of its own information symbols. In phase one, receiver two observes $N_2(N_2 - M_1)$ independent equations in $\sum_{t=1}^{W_1} \omega_t + \eta_1 = (N_2 - M_1)(N_1 + N_2)$ unknowns. Including $N_1(N_2 - M_1)$ independent linear combinations obtained from phase two, receiver two will have $N_2(N_2 - M_1) + N_1(N_2 - M_1) = (N_2 - M_1)(N_1 + N_2)$ equations in $(N_2 - M_1)(N_1 + N_2)$ unknowns. Therefore, receiver two will be able to resolve its desired information symbols. This completes the proof of achievability for corner point $T_9$.

## VI. Proof of Theorem 4

In this section, we prove that our achievable DoF region is tight for the antenna configurations stated in Theorem 4. To show this, we need an outer-bound for the DoF region of the two-user MIMO Gaussian IC with delayed local CSIT. The following two regions can be served as an outer-bound for the DoF

region of this channel:

1) The DoF region of the two-user MIMO Gaussian IC with perfect CSIT which has been characterized in [10]. This region which will be denoted by $\mathscr{D}_{\text{IC}}^{\text{p-CSI}}$ is the union of all $(d_1, d_2) \in \mathbb{R}_+^2$ that satisfy the following three inequalities [10]

$$\begin{aligned} d_1 &\leq \min(M_1, N_1) \\ d_2 &\leq \min(M_2, N_2) \\ d_1 + d_2 &\leq \min\{M_1 + M_2, N_1 + N_2, \max(M_1, N_2), \max(M_2, N_1)\}. \end{aligned} \tag{13}$$

2) The DoF region of the two-user MIMO broadcast channel (BC) with delayed CSIT which has been recently characterized in [9]. Let $\mathscr{D}_{\text{BC}}^{\text{d-CSI}}$ denote the DoF region of a two-user MIMO broadcast channel with $M$ antennas at the transmitter and $N_1, N_2$ antennas at the receivers. In [9], the authors proved that the region $\mathscr{D}_{\text{BC}}^{\text{d-CSI}}$ is the union of all $(d_1, d_2) \in \mathbb{R}_+^2$ that satisfy the following two inequalities:

$$\begin{aligned} \tfrac{d_1}{\min(M, N_1+N_2)} + \tfrac{d_2}{\min(M, N_2)} &\leq 1 \\ \tfrac{d_1}{\min(M, N_1)} + \tfrac{d_2}{\min(M, N_1+N_2)} &\leq 1. \end{aligned} \tag{14}$$

The intersection of the above two outer-bounds is again an outer-bound. We will compare our achievable DoF region with this outer-bound in the following:

**Class** $\mathcal{C}_1$: $M_2 \leq N_1$

The DoF region with no CSIT and perfect CSIT are identical in this case and are equal to $\mathscr{D}_{\text{IC,in}}^{\text{d-CSI}}$ [3], [4]. Therefore, $\mathscr{D}_{\text{IC}}^{\text{d-CSI}} = \mathscr{D}_{\text{IC,in}}^{\text{d-CSI}}$. The broadcast outer-bound is larger than the full CSIT outer-bound in this case. Both outer-bounds as well as our achievable DoF region are depicted in Fig. 4(a).

**Class** $\mathcal{C}_2$: $\min(M_1, M_2) > N_1$ and $M_2 \leq N_2$

As we can see in Fig. 4(b), the intersection of broadcast outer-bound and perfect CSIT outer-bound is tighter than each of them in this class. Our achievable DoF region is strictly smaller than this outer-bound. There is no specific configuration in this class that our achievable region coincide with the outer-bound region.

**Class** $\mathcal{C}_3$ : $N_1 < M_1 \leq N_2 < M_2$

For this class, the intersection of broadcast outer-bound and full CSIT outer-bound is tighter than each of them and is described by $\frac{d_1}{N_1} + \frac{d_2}{N_1+N_2} \leq 1$ and $d_1 + d_2 \leq N_2$. On the other hand, the region $\mathscr{D}_{\text{IC,in}}^{\text{d-CSI}}$ is described by $d_1 + d_2 \leq N_2$ and $\frac{d_1}{N_1} + \frac{d_2}{M_2'} \leq 1$ in this class. Therefore, the intersection of the broadcast outer-bound and perfect CSIT outer-bound coincides with our achievable DoF region provided that $M_2 \geq N_1 + N_2$. Notice that for this special case, $\mathscr{D}_{\text{IC}}^{\text{d-CSI}}$ is strictly larger than the DoF region with no CSIT and is strictly smaller than DoF region with perfect CSIT. The outer-bounds and our achievable

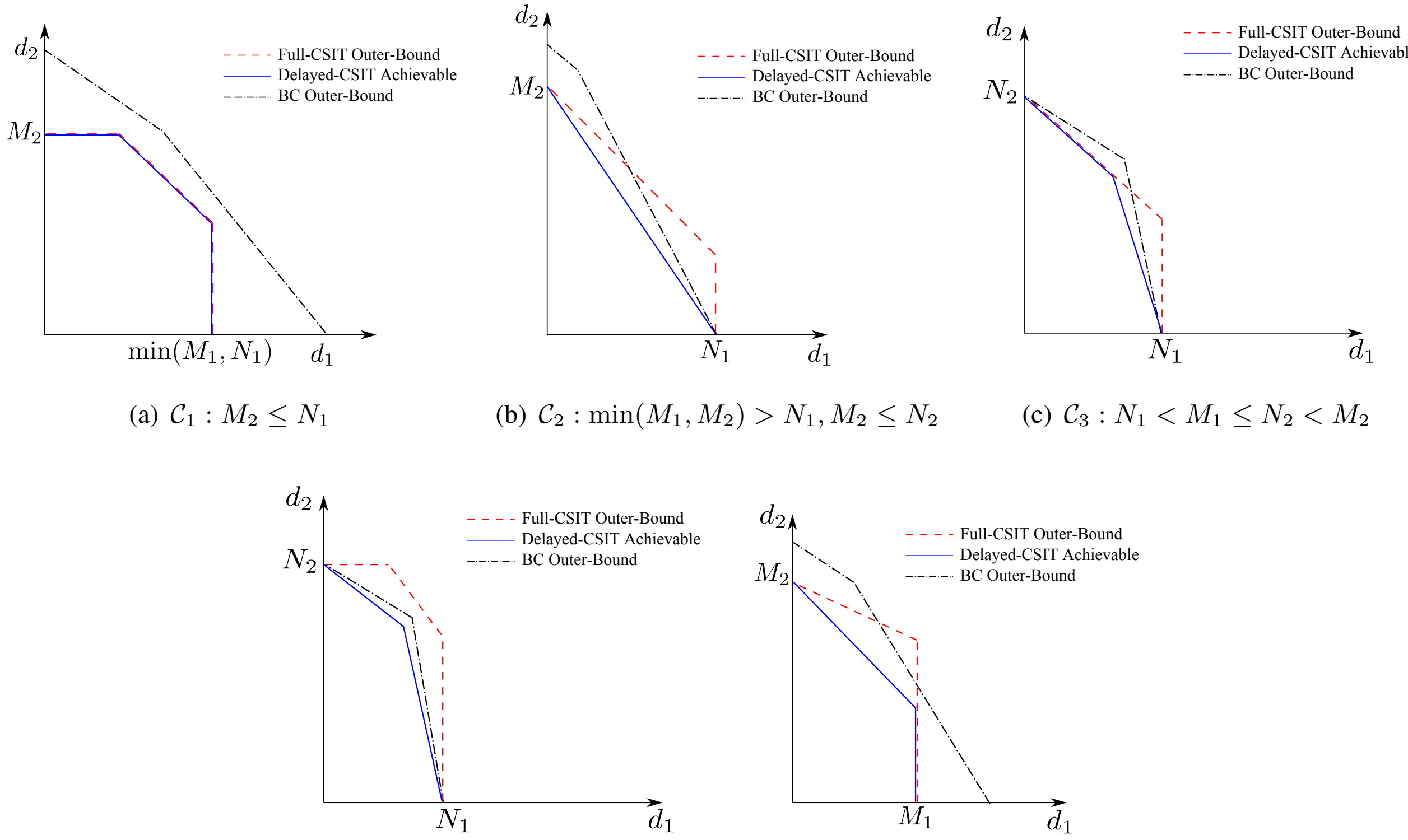

(a) $\mathcal{C}_1 : M_2 \leq N_1$ (b) $\mathcal{C}_2 : \min(M_1, M_2) > N_1, M_2 \leq N_2$ (c) $\mathcal{C}_3 : N_1 < M_1 \leq N_2 < M_2$

(d) $\mathcal{C}_4 : \min(M_1, M_2) > N_2 \geq N_1$ (e) $\mathcal{C}_5 : M_1 \leq N_1 < M_2 \leq N_2$

Fig. 4. Outer-bounds on the DoF region of the two-user MIMO Gaussian IC with $N_2 \geq N_1$ and with delayed CSIT for different classes. Our achievable DoF region is also presented for comparison.

region for this class are depicted in Fig. 4(c).

**Class $\mathcal{C}_4 : \min(M_1, M_2) > N_2 \geq N_1$**

For this class, the broadcast outer-bound is tighter than full CSIT outer-bound and is described by $\frac{d_1}{N_1+N_2} + \frac{d_2}{N_2} \leq 1$ and $\frac{d_1}{N_1} + \frac{d_2}{N_1+N_2} \leq 1$. The region $\mathscr{D}_{\text{IC,in}}^{\text{d-CSI}}$ is described by $\frac{d_1}{M'1} + \frac{d_2}{N_2} \leq 1$ and $\frac{d_1}{N_1} + \frac{d_2}{M_2'} \leq 1$. Therefore, if $\min(M_1, M_2) \geq N_1 + N_2$, our achievable region will be tight. Again for this special case $\mathscr{D}_{\text{IC}}^{\text{d-CSI}}$ is strictly larger than the DoF region with no CSIT and is strictly smaller than DoF region with perfect CSIT. The outer-bounds and our achievable region for this class are depicted in FIg. 4(d).

**Class $\mathcal{C}_5 : M_1 \leq N_1 < M_2 \leq N_2$**

As we can see in Fig. 4(e), the intersection of broadcast outer-bound and perfect CSIT outer-bound is tighter than each of them in this class and is described by $d_1 \leq M_1$, $d_1 + d_2 \leq M_2$, and $\frac{d_1}{N_1} + \frac{d_2}{M_1+M_2} \leq 1$. Our achievable DoF region is described by $d_1 \leq M_1$ and $\frac{d_1}{N_1} + \frac{d_2}{M_2} \leq 1$. Therefore, there is no specific configuration in this class for that our achievable region $\mathscr{D}_{\text{IC,in}}^{\text{p-CSI}}$ coincides with the outer-bound.

**Class $\mathcal{C}_6 : M_1 \leq N_1 \leq N_2 < M_2$**

For subclass $\mathcal{C}_{61}$ and $\mathcal{C}_{62}$, the perfect CSIT outer bound is tighter than the broadcast outer-bound and coincides with our achievable DoF region. This case is depicted in Fig. 5(a). For subclass $\mathcal{C}_{63}$, the intersection of broadcast outer-bound and perfect CSIT outer-bound is tighter than each of them and

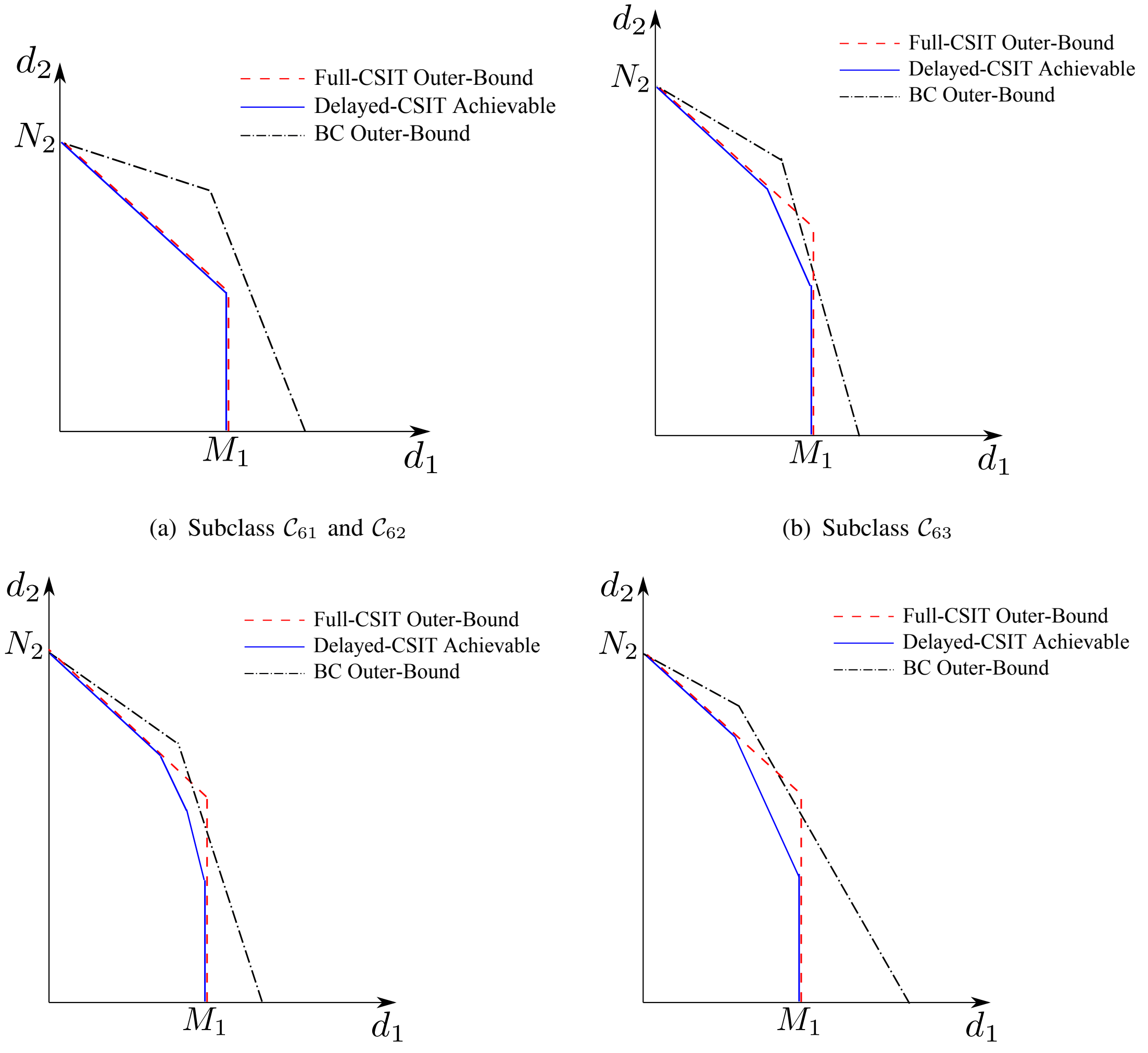

(a) Subclass $\mathcal{C}_{61}$ and $\mathcal{C}_{62}$

(b) Subclass $\mathcal{C}_{63}$

(c) Subclass $\mathcal{S}_1$

(d) Subclass $\mathcal{S}_2$

Fig. 5. Outer-bounds on the DoF region of the two-user MIMO Gaussian IC with $N_2 \geq N_1$ and with delayed-CSIT for class $\mathcal{C}_6$ ($M_1 \leq N_1 \leq N_2 < M_2$). Our achievable DoF region is also presented for comparison.

is described by $d_1 \leq M_1$, $d_1 + d_2 \leq N_2$, and $\frac{d_1}{N_1} + \frac{d_2}{M_1+M_2} \leq 1$. Our achievable DoF region for this subclass is described by $d_1 \leq M_1$, $d_1 + d_2 \leq N_2$, and $\frac{d_1}{N_1} + \frac{d_1}{M_2} \leq 1$. Therefore, there is no specific configuration in this subclass for which our achievable DoF region is tight. This case is depicted in Fig. 5(b). For subclass $\mathcal{S}_1$, the intersection of broadcast outer-bound and perfect CSIT outer-bound is tighter than each of them and is described by $d_1 \leq M_1$, $d_1 + d_2 \leq N_2$, and $\frac{d_1}{N_1} + \frac{d_2}{N_1+N_2} \leq 1$. Our achievable DoF region is strictly smaller than this outer-bound as depicted in Fig. 5(c). For subclass $\mathcal{S}_2$, the intersection of the tow outer-bounds is generally tighter than each of them and is described by $d_1 \leq M_1$, $d_1 + d_2 \leq N_2$, and $\frac{d_1}{N_1} + \frac{d_2}{N_1+N_2} \leq 1$. Our achievable DoF region for this subclass is described by the corner points $(0, 0)$, $(M_1, 0)$, $(0, N_2)$, $(\frac{N_1^2}{N_1+N_2-M_1}, N_2 - \frac{N_1^2}{N_1+N_2-M_1})$, and $(M_1, \frac{(N_1+N_2-M_1)(N_1-M_1)}{N_1})$. As we can see in Fig. 5(d), our achievable DoF region does not coincide with the outer-bound for this class.

## VII. Proof of Theorem 5

In this section we prove Theorem 5. We first show that for a $K$-user MISO Gaussian IC with $M \geq K$ antennas at each transmitter and with delayed local CSIT, the sum-DoF is upper-bounded by $\frac{K^2-2K+2}{K^2-K+1}K$. We then show that with delayed local CSIT, we can achieve a sum-DoF of $\frac{K}{K^2-K+1}K$ for this channel. Consider a $K$-user MISO Gaussian IC with $M \geq K$ antennas at each transmitter and with delayed local CSIT. If we allow full cooperation among $K-1$ transmitters and full cooperation among their corresponding receivers, we reach to a two-user MIMO IC with $\tilde{M}_1 = (K-1)M$ and $\tilde{M}_2 = M$ antennas at the transmitters and $\tilde{N}_1 = K-1$ and $\tilde{N}_2 = 1$ at their corresponding receivers. Since cooperation can not reduce capacity, the sum-DoF of the resulting two-user MIMO channel with delayed local CSIT will be an upper-bound for the sum-DoF of the original $K$ users MISO IC with delayed local CSIT. The resulting two-user MIMO IC falls into the class $\mathcal{C}_4$ and moreover $\min(\tilde{M}_1, \tilde{M}_2) \geq \tilde{N}_1 + \tilde{N}_2$. Hence, from the results in section II, one can write

$$d_1 + d_1 + \cdots + d_K \leq (1 - \frac{\tilde{N}_1\tilde{N}_2}{\tilde{N}_1^2 + \tilde{N}_2^2 + \tilde{N}_1\tilde{N}_2})(\tilde{N}_1 + \tilde{N}_2) = \frac{K^2-2K+2}{K^2-K+1}K,$$

which is the desired result. With full CSIT, the sum-DoF of the $K$ user MISO IC with $M$ antennas at each transmitter is equal to $K$. For example, for a three-user MISO IC with $M = 5$, the sum-DoF with full CSIT is equal to $3$ while the sum-DoF with delayed local CSIT is upper-bounded by $\frac{15}{7} \approxeq 2.14$. Now, we prove that we can achieve a sum-DoF of $\frac{K}{K^2-K+1}K$ for the MISO IC with $M \geq K$ antennas at each transmitter and with delayed local CSIT. Our transmission scheme consists of two phases:

Phase One: This phase takes one time slot. In this phase, each transmitter sends $K$ independent coded information symbols intended for its corresponding receiver. Since $M \geq K$, each transmitter employs only $K$ out of its $M$ transmit antennas and nothing is sent on the remaining $M-K$ transmit antennas.

Phase Two: This phase takes $K(K-1)$ time slots. No new information is sent during this phase. The first $K-1$ time slots of this phase is dedicated to transmitter one, the next $K-1$ time slots is dedicated to transmitter two and so on. During the first $K-1$ time slots of phase two, only the the first transmitter is active and all other transmitters are silent. At each time slot of these $K-1$ time slots, transmitter one sends one of those $K-1$ interference terms that it has generated in its non-intended receivers during the phase one. It is important to notice that interference terms that transmitter one generates in its non-intended receivers during the phase one are functions of information symbols of this transmitter and local channel CSI at this transmitter and therefore, transmitter one has access to them at the consequent time slots. Similarly, at the next $K-1$ time slots, only transmitter two is active and it sends those $K-1$ interference terms that it has generated in its non-intended receivers during the phase one. It is easy to see that after $K(K-1)$ time slots, each receiver is able to resolve all its intended information symbols.

Since $K^2$ information symbols have been transmitted in $1+K(K-1)$ time slots, we achieve a sum-DoF of $\frac{K^2}{K^2-K+1}$. This completes the proof.

## VIII. Conclusion

We obtained new results for the DoF region of the two-user MIMO Gaussian IC with delayed local CSIT. We showed how interference alignment technique can be applied to this channel to achieve a larger DoF region. We also compared our achievable DoF region under delayed local CSIT assumption with DoF region with no CSIT and perfect CSIT. By employing some simple outer-bounds, we proved that our achievable DoF region is tight in some cases. We also provided lower and upper-bounds on the sum-DoF of the $K$-user MISO Gaussian IC with delayed local CSIT and with $M \geq K$.

## Appendix A
## Derivation of rank of matrix $\mathbf{P}^{[i]}$

Define $\mathbf{diag}(\{\mathbf{A}_k\}_{k=1}^n)$, where $\{\mathbf{A}_k\}_{k=1}^n$ is a sequence of $n$ arbitrary matrices, as follows:

$$\mathbf{diag}(\{\mathbf{A}_k\}_{k=1}^n) = \begin{pmatrix} \mathbf{A}_1 & & \bigcirc \\ & \mathbf{A}_2 & \\ & & \ddots & \\ \bigcirc & & & \mathbf{A}_n \end{pmatrix}.$$

Also let $\mathbf{H}_{ij}(t)$ denote the $N_i \times M'_j$ matrix of the channel coefficients between transmitter $j$ and receiver $i$ at time slot $t$. During the phase one, transmitter $\bar{i}$ generates the interference vector $\mathbf{diag}(\{\mathbf{H}_{i\bar{i}}(t)\}_{t=1}^{W_i})\mathbf{U}^{[\bar{i}]}$ at receiver $i$. The matrix $\mathbf{diag}(\{\mathbf{H}_{i\bar{i}}(t)\}_{t=1}^{W_i})$ is a random $N_iW_i \times M'_{\bar{i}}W_i$ matrix whose rank can be easily shown to be $\min(M'_{\bar{i}}, N_i)W_i$, almost surely. Therefore, it can be decomposed as $\mathbf{diag}(\{\mathbf{H}_{i\bar{i}}(t)\}_{t=1}^{W_i} = \mathbf{H}'_{i\bar{i}}\mathbf{H}''_{i\bar{i}}$, where $\mathbf{H}'_{i\bar{i}}$ and $\mathbf{H}''_{i\bar{i}}$ are of size $N_iW_i \times \min(M'_{\bar{i}}, N_i)W_i$ and $\min(M'_{\bar{i}}, N_i)W_i \times M'_{\bar{i}}W_i$, respectively, and both have rank $\min(M'_{\bar{i}}, N_i)W_i$. Such a decomposition is trivial because one of $\mathbf{H}'_{i\bar{i}}$ and $\mathbf{H}''_{i\bar{i}}$ is the identity matrix and the other is $\mathbf{diag}(\{\mathbf{H}_{i\bar{i}}(t)\}_{t=1}^{W_i})$. We define $\mathbf{I}^{[i]} \triangleq \mathbf{H}''_{i\bar{i}}\mathbf{U}^{[\bar{i}]}$. Hence, the interference observed by receiver $i$ is obtained by $\mathbf{H}'_{i\bar{i}}\mathbf{I}^{[i]}$. The vector $\mathbf{I}^{[i]}$ is of size $\min(M'_{\bar{i}}, N_i)W_{\bar{i}}$ and is called the *effective interference vector* at receiver $i$. After $W$ channel uses, receiver $i$ has $N_iW$ linear equations in terms of the elements of the vector $\mathbf{e}_i \triangleq [\mathbf{U}^{[i]}, \mathbf{I}^{[i]}]$ where $\mathbf{U}^{[i]} = [u_1^{[i]}, \cdots, u_{M'_iW_i}^{[i]}]^T$ and $\mathbf{I}^{[i]}$ is the effective interference vector at receiver $i$. Let $\mathbf{P}^{[i]}$ denote the coefficient matrix of this set of linear equations for receiver $i$. So, $\mathbf{P}^{[i]}$ is of size $N_iW \times [M'_iW_i + \min(M'_{\bar{i}}, N_i)W_{\bar{i}}]$. Based on the our transmission scheme, the matrix $\mathbf{P}^{[i]}$ can be partitioned into six random sub-matrices $\mathbf{P}_{11}^{[i]}, \mathbf{P}_{21}^{[i]}, \mathbf{P}_{31}^{[i]}, \mathbf{P}_{12}^{[i]}, \mathbf{P}_{22}^{[i]}, \mathbf{P}_{32}^{[i]}$ as follows:

$$\mathbf{P}^{[i]} = \left(\begin{array}{c|c} \mathbf{P}_{11}^{[i]} & \mathbf{P}_{12}^{[i]} \\ \hline \mathbf{P}_{21}^{[i]} & \mathbf{P}_{22}^{[i]} \\ \hline \mathbf{P}_{31}^{[i]} & \mathbf{P}_{32}^{[i]} \end{array}\right)$$

Based on the values of $W_i$ and $W_{\bar{i}}$, we consider two different cases:

(a) $W_i \leq W_{\bar{i}}$

- $\mathbf{P}_{11}^{[i]}$ is the coefficient submatrix of size $N_i W_i \times M_i' W_i$ whose elements are the coefficients of the information symbols generated by transmitter $i$ during the first $W_i$ time slots. It is easily seen that this matrix is indeed equal to $\mathbf{diag}(\{\mathbf{H}_{ii}(t)\}_{t=1}^{W_i})$. So, $\mathbf{P}_{11}^{[i]}$ is a random matrix of rank $\min(M_i', N_i)W_i$, almost surely.
- $\mathbf{P}_{21}^{[i]}$ and $\mathbf{P}_{31}^{[i]}$ are the coefficient sub-matrices of size $N_i(W_{\bar{i}} - W_i) \times M_i' W_i$ and $N_i(W - W_{\bar{i}}) \times M_i' W_i$, respectively and are given by the following matrix multiplications

$$\mathbf{P}_{21}^{[i]} = \mathbf{diag}(\{\mathbf{H}_{ii}(t)\}_{t=W_i+1}^{W_{\bar{i}}})\mathbf{G}_i \mathbf{H}_{\bar{i}i}' \tag{15}$$

$$\mathbf{P}_{31}^{[i]} = \mathbf{diag}(\{\mathbf{H}_{ii}(t)\}_{t=W_{\bar{i}}+1}^{W})\mathbf{G}_i' \mathbf{H}_{\bar{i}i}', \tag{16}$$

where $\mathbf{G}_i$ and $\mathbf{G}_i'$ are the matrices of size $M_i'(W_{\bar{i}} - W_i) \times \min(M_i', N_{\bar{i}})W_i$ and $M_i'(W - W_{\bar{i}}) \times \min(M_i', N_{\bar{i}})W_i$, respectively, containing the random coefficients used by transmitter $i$ to send the random linear combinations of the effective interference quantities at receiver $\bar{i}$ during the last $W - W_i$ channel uses. Therefore, $\mathbf{G}_i$ (resp. $\mathbf{G}_i'$) is of rank $\min\{M_i'(W_{\bar{i}} - W_i), \min(M_i', N_{\bar{i}})W_i\}$ (resp. $\min\{M_i'(W - W_{\bar{i}}), \min(M_i', N_{\bar{i}})W_i\}$), almost surely.

Since the above matrices are random matrices generated independent of each other, the rank of their multiplication would be the minimum value of their ranks, almost surely. Hence,

$$\begin{aligned} \text{rank}(\mathbf{P}_{21}^{[i]}) &= \min\{\text{rank}(\mathbf{diag}(\{\mathbf{H}_{ii}(t)\}_{t=W_i+1}^{W_{\bar{i}}})), \text{rank}(\mathbf{G}_i), \text{rank}(\mathbf{H}_{\bar{i}i}')\} \\ &= \min\{\min(M_i', N_i)(W_{\bar{i}} - W_i), \min\{M_i'(W_{\bar{i}} - W_i), \min(M_i', N_{\bar{i}})W_i\}, \min(M_i', N_{\bar{i}})W_i\} \\ &= \min\{\min(M_i', N_i)(W_{\bar{i}} - W_i), \min(M_i', N_{\bar{i}})W_i\}, \end{aligned} \tag{17}$$

and Similarly,

$$\text{rank}(\mathbf{P}_{31}^{[i]}) = \min\{\min(M_i', N_i)(W - W_{\bar{i}}), \min(M_i', N_{\bar{i}})W_i\}. \tag{18}$$

- $\mathbf{P}_{12}^{[i]}$ and $\mathbf{P}_{22}^{[i]}$ are the coefficient sub-matrices of size $N_i W_i \times \min(M_{\bar{i}}', N_i)W_{\bar{i}}$ and $N_i(W_{\bar{i}} - W_i) \times \min(M_{\bar{i}}', N_i)W_{\bar{i}}$, respectively, whose elements are the coefficients used by transmitter $\bar{i}$ to send its own information symbols during the first $W_{\bar{i}}$ channel uses. It is easily seen that $\mathbf{P}_{12}^{[i]}$ (resp. $\mathbf{P}_{22}^{[i]}$) is composed of the first $N_i W_i$ rows (resp. last $N_i(W_{\bar{i}} - W_i)$ rows) of $\mathbf{H}_{i\bar{i}}'$. So, $\text{rank}(\mathbf{P}_{12}^{[i]}) = \min(M_{\bar{i}}', N_i)W_i$ and $\text{rank}(\mathbf{P}_{22}^{[i]}) = \min(M_{\bar{i}}', N_i)(W_{\bar{i}} - W_i)$, almost surely.
- $\mathbf{P}_{32}^{[i]}$ is the coefficient sub-matrix of size $N_i(W - W_{\bar{i}}) \times \min(M_{\bar{i}}', N_i)W_{\bar{i}}$ whose elements are the coefficients used by transmitter $\bar{i}$ to send the effective interference quantities at receiver $i$ during the last $W - W_{\bar{i}}$ channel uses. This matrix is obtained by the following matrix multiplication

$$\mathbf{P}_{32}^{[i]} = \textbf{diag}(\{\mathbf{H}_{i\bar{i}}(t)\}_{t=W_{\bar{i}}+1}^{W})\mathbf{D}_{\bar{i}}. \tag{19}$$

The matrix $\mathbf{D}_{\bar{i}}$ is a matrix of size $M_{\bar{i}}'(W - W_{\bar{i}}) \times \min(M_{\bar{i}}', N_i)W_{\bar{i}}\}$ containing the random coefficients used by transmitter $\bar{i}$ to send random linear combinations of the effective interference quantities at receiver $i$ during the last $W - W_{\bar{i}}$ channel uses. Matrices $\textbf{diag}(\{\mathbf{H}_{i\bar{i}}(t)\}_{t=W_{\bar{i}}+1}^{W})$ and $\mathbf{D}_{\bar{i}}$ are random matrices generated independent of each other, and thus, the rank of their multiplication would be the minimum value of their ranks, almost surely. Therefore,

$$\begin{aligned}
\text{rank}(\mathbf{P}_{32}^{[i]}) &= \min\{\text{rank}(\textbf{diag}(\{\mathbf{H}_{\bar{i}i}(k)\}_{k=W_{\bar{i}}+1}^{W})), \text{rank}(\mathbf{D}_{\bar{i}})\} \\
&= \min\{\min(M_{\bar{i}}', N_i)(W - W_{\bar{i}}), \min\{M_{\bar{i}}'(W - W_{\bar{i}}), \min(M_{\bar{i}}', N_i)W_{\bar{i}}\}\} \\
&= \min\{\min(M_{\bar{i}}', N_i)(W - W_{\bar{i}}), \min(M_{\bar{i}}', N_i)W_{\bar{i}}\} \\
&= \min(M_{\bar{i}}', N_i)\min(W - W_{\bar{i}}, W_{\bar{i}}).
\end{aligned} \tag{20}$$

(b) $W_i > W_{\bar{i}}$

Parallel to the result for case $W_i \leq W_{\bar{i}}$, we have the following results for this case:

$$\begin{aligned}
\text{rank}(\mathbf{P}_{11}^{[i]}) &= \min(M_i', N_i)W_{\bar{i}} \\
\text{rank}(\mathbf{P}_{21}^{[i]}) &= \min(M_i', N_i)(W_i - W_{\bar{i}}) \\
\text{rank}(\mathbf{P}_{31}^{[i]}) &= \min\{\text{rank}(\textbf{diag}(\{\mathbf{H}_{ii}(t)\}_{t=W_i+1}^{W})), \text{rank}(\mathbf{D}_i), \text{rank}(\mathbf{H}_{\bar{i}i}')\} \\
&= \min\{\min(M_i', N_i)(W - W_i), \min\{M_i'(W - W_i), \min(M_i', N_{\bar{i}})W_i\}, \min(M_i', N_{\bar{i}})W_i\} \\
&= \min\{\min(M_i', N_i)(W - W_i), \min(M_i', N_{\bar{i}})W_i\} \\
\text{rank}(\mathbf{P}_{12}^{[i]}) &= \min(M_{\bar{i}}', N_i)W_{\bar{i}} \\
\text{rank}(\mathbf{P}_{22}^{[i]}) &= \min\{\text{rank}(\textbf{diag}(\{\mathbf{H}_{i\bar{i}}(k)\}_{k=W_{\bar{i}}+1}^{W_i})), \text{rank}(\mathbf{G}_{\bar{i}})\} \\
&= \min\{\min(M_{\bar{i}}', N_i)(W_i - W_{\bar{i}}), \min\{M_{\bar{i}}'(W_i - W_{\bar{i}}), \min(M_{\bar{i}}', N_i)W_{\bar{i}}\}\} \\
&= \min\{\min(M_{\bar{i}}', N_i)(W_i - W_{\bar{i}}), \min(M_{\bar{i}}', N_i)W_{\bar{i}}\} \\
&= \min(M_{\bar{i}}', N_i)\min(W_i - W_{\bar{i}}, W_{\bar{i}}) \\
\text{rank}(\mathbf{P}_{32}^{[i]}) &= \min(M_{\bar{i}}', N_i)\min(W - W_i, W_{\bar{i}}).
\end{aligned}$$

Now, by horizontal concatenation of $\mathbf{P}_{m1}^{[i]}$ and $\mathbf{P}_{m2}^{[i]}$, $m = 1, 2, 3$, one can write

(a) $W_i \leq W_{\bar{i}}$

$$\begin{aligned}
r_1^{[i]} &= \min(N_iW_i, \min(M_i', N_i)W_i + \min(M_{\bar{i}}', N_i)W_i) \\
&= \min(N_iW_i, M_i'W_i + M_{\bar{i}}'W_i) \\
&= \min(N_i, M_i' + M_{\bar{i}}')W_i \\
r_2^{[i]} &= \min\{N_i(W_{\bar{i}} - W_i), \min\{\min(M_i', N_i)(W_{\bar{i}} - W_i), \min(M_i', N_{\bar{i}})W_i\} + \min(M_{\bar{i}}', N_i)(W_{\bar{i}} - W_i)\} \\
&= \min\{N_i(W_{\bar{i}} - W_i), \min\{M_i'(W_{\bar{i}} - W_i), M_i'W_i, N_{\bar{i}}W_i\} + M_{\bar{i}}'(W_{\bar{i}} - W_i)\} \\
r_3^{[i]} &= \min\{N_i(W - W_{\bar{i}}), \min\{\min(M_i', N_i)(W - W_{\bar{i}}), \min(M_i', N_{\bar{i}})W_i\} + \min(M_{\bar{i}}', N_i)\min(W - W_{\bar{i}}, W_{\bar{i}})\} \\
&= \min\{N_i(W - W_{\bar{i}}), \min\{M_i'(W - W_{\bar{i}}), M_i'W_i, N_{\bar{i}}W_i\} + \min\{M_{\bar{i}}'(W - W_{\bar{i}}), M_{\bar{i}}'W_{\bar{i}}, N_iW_{\bar{i}}\}\}
\end{aligned}$$

(b) $W_i > W_{\bar{i}}$

$$\begin{aligned}
r_1^{[i]} &= \min(N_i W_{\bar{i}}, \min(M_i', N_i) W_{\bar{i}} + \min(M_{\bar{i}}', N_i) W_{\bar{i}}) \\
&= \min(N_i, M_i' + M_{\bar{i}}') W_{\bar{i}} \\
r_2^{[i]} &= \min\{N_i(W_i - W_{\bar{i}}), \min(M_i', N_i)(W_i - W_{\bar{i}}) + \min(M_{\bar{i}}', N_i) \min(W_i - W_{\bar{i}}, W_{\bar{i}})\} \\
&= \min\{N_i(W_i - W_{\bar{i}}), M_i'(W_i - W_{\bar{i}}) + \min\{M_{\bar{i}}'(W_i - W_{\bar{i}}), M_{\bar{i}}' W_{\bar{i}}, N_i W_{\bar{i}}\}\} \\
r_3^{[i]} &= \min\{N_i(W - W_i), \min\{\min(M_i', N_i)(W - W_i), \min(M_i', N_{\bar{i}}) W_i\} + \min(M_{\bar{i}}', N_i) \min(W - W_i, W_{\bar{i}})\} \\
&= \min\left\{N_i(W - W_i), \min\{M_i'(W - W_i), M_i' W_i, N_{\bar{i}} W_i\} + \min\{M_{\bar{i}}'(W - W_i), M_{\bar{i}}' W_{\bar{i}}, N_i W_{\bar{i}}\}\right\}
\end{aligned}$$

where, $r_m^{[i]}$ is the rank of the horizontal concatenation of $\mathbf{P}_{m1}^{[i]}$ and $\mathbf{P}_{m2}^{[i]}$, $m = 1, 2, 3$. Now, $\mathbf{P}^{[i]}$ is obtained by vertical concatenation of the three resulting matrices.

The above results can be summarized as follows: Define $i_{\text{max}} \triangleq \arg\max_i \{W_i\}$ and $i_{\text{min}} \triangleq \{1, 2\} \backslash \{i_{\text{max}}\}$. Then, one can write

$$\begin{aligned}
r_1^{[i]} &= \min\left\{N_i, M_i' + M_{\bar{i}}'\right\} W_{i_{\text{min}}} \\
r_2^{[i]} &= \min\left\{N_i(W_{i_{\text{max}}} - W_{i_{\text{min}}}), M_{i_{\text{max}}}'(W_{i_{\text{max}}} - W_{i_{\text{min}}}) + \min\left( M_{i_{\text{min}}}'(W_{i_{\text{max}}} - W_{i_{\text{min}}}), M_{i_{\text{min}}}' W_{i_{\text{min}}}, N_{i_{\text{max}}} W_{i_{\text{min}}}\right)\right\} \\
r_3^{[i]} &= \min\left\{N_i(W - W_{i_{\text{max}}}), \min\left( M_i'(W - W_{i_{\text{max}}}), M_i' W_i, N_{\bar{i}} W_i\right) + \min\left( M_{\bar{i}}'(W - W_{i_{\text{max}}}), M_{\bar{i}}' W_{\bar{i}}, N_i W_{\bar{i}}\right)\right\}.
\end{aligned}$$

Therefore, rank of $\mathbf{P}^{[i]}$ can be computed as follows:

$$\text{rank}(\mathbf{P}^{[i]}) = \min\{M_i' W_i + \min(M_{\bar{i}}', N_i) W_{\bar{i}}, r_1^{[i]} + r_2^{[i]} + r_3^{[i]}\},$$

which is the desired result.

## APPENDIX B
## PROOF OF ACHIEVABILITY FOR CLASS $\mathcal{C}_3$

To show point $Q_3 = (\frac{N_1(M_2'-N_2)}{M_2'-N_1}, \frac{M_2'(N_2-N_1)}{M_2'-N_1})$ is achievable, we show that $W = M_1(M_2' - N_1)$, $W_1 = N_1(M_2' - N_2)$, and $W_2 = M_1(N_2 - N_1)$ satisfy the rank conditions in (9). First, we notice that $W = \frac{M_1}{N_1} W_1 + W_2$ and hence $W > W_1 + W_2$. We need to differentiate between two cases:

- $W_1 \geq W_2$

From $W_1 \geq W_2$, it follows that $M_1 < \frac{N_1(M_2'-N_2)}{N_2-N_1}$ and we have the following chain of inequalities

$$N_1 < M_1 < \frac{N_1(M_2' - N_2)}{N_2 - N_1} \overset{(a)}{\leq} \frac{N_1^2}{N_2 - N_1} \Rightarrow N_2 \leq 2N_1, \tag{21}$$

where (a) follows from $M_2' \leq N_1 + N_2$. Moreover, we have

$$\frac{W}{W_1} \overset{(a)}{\leq} \frac{M_1 N_2}{W_1} \overset{(b)}{\leq} \frac{M_1 N_2}{W_2} = \frac{N_2}{N_2 - N_1} \Rightarrow N_1 W > N_2(W - W_1), \tag{22}$$

where (a) follows from $M_2' \leq N_1 + N_2$ and (b) follows from $W_1 \geq W_2$. Substituting $i_{\text{max}} = 1$ and $i_{\text{min}} = 2$ in (9), we have

$$\begin{aligned} r_1^{[1]} &= \min\{N_1, M_1 + M_2'\} W_2 \overset{(a)}{=} N_1 W_2 \\ r_2^{[1]} &= \min\Big\{N_1(W_1 - W_2), M_1(W_1 - W_2) + \min\Big( M_2'(W_1 - W_2), M_2' W_2, N_1 W_2\Big)\Big\} \overset{(b)}{=} N_1(W_1 - W_2) \\ r_3^{[1]} &= \min\Big\{N_1(W - W_1), \min\Big( M_1(W - W_1), M_1 W_1, N_2 W_1\Big) + \min\Big( M_2'(W - W_1), M_2' W_2, N_1 W_2\Big)\Big\} \\ &\overset{(c)}{=} \min\Big\{N_1(W - W_1), \min\Big( M_1(W - W_1), M_1 W_1\Big) + N_1 W_2\Big\} \overset{(d)}{=} N_1(W - W_1), \end{aligned}$$

where (a), (b), and (c) follow from the assumptions $N_1 < M_1 \leq N_2 < M_2' \leq N_1 + N_2$ and $W - W_1 > W_2$ and (d) follows from $N_1 < M_1$ and the fact that $M_1 W_1 + N_1 W_2 = N_1 W$. Therefore, $r_1^{[1]} + r_2^{[1]} + r_3^{[1]} = N_1 W$. On the other hand, $M_1 W_1 + \min(M_2', N_1) W_2 = M_1 W_1 + N_1 W_2 = N_1 W$ and hence the rank condition is met for receiver one. For receiver two, we have

$$\begin{aligned} r_1^{[2]} &= \min\{N_2, M_1 + M_2'\} W_2 \overset{(a)}{=} N_2 W_2 \\ r_2^{[2]} &= \min\Big\{N_2(W_1 - W_2), M_1(W_1 - W_2) + \min\Big( M_2'(W_1 - W_2), M_2' W_2, N_1 W_2\Big)\Big\} \overset{(b)}{=} N_2(W_1 - W_2) \\ r_3^{[2]} &= \min\Big\{N_2(W - W_1), \min\Big( M_2'(W - W_1), M_2' W_2, N_1 W_2\Big) + \min\Big( M_1(W - W_1), M_1 W_1, N_2 W_1\Big)\Big\} \\ &\overset{(c)}{=} \min\Big\{N_2(W - W_1), N_1 W_2 + \min\Big( M_1(W - W_1), M_1 W_1\Big)\Big\} \overset{(d)}{=} N_2(W - W_1), \end{aligned}$$

where (a) follows from the assumption $N_2 < M_2'$, (b) follows from the assumption $N_1 < N_2 < M_2'$ and the inequality $M_1(W_1 - W_2) + N_1 W_2 > N_2(W_1 - W_2)$ which is valid because of the following chain of inequalities $\frac{W_1}{W_2} = \frac{N_1}{M_1} \frac{M_2' - N_2}{N_2 - N_1} < \frac{M_2' - N_2}{N_2 - N_1} \leq \frac{N_1}{N_2 - M_1}$, (c) follows from $N_1 < M_1 < M_2'$, $W - W_1 > W_2$, and (d) follows from the inequalities $N_1 W_2 + M_1(W - W_1) > N_2(W - W_1)$ and $N_1 W_2 + M_1 W_1 > N_2(W - W_1)$ which are direct consequences of (21) and (22). Therefore, $r_1^{[2]} + r_2^{[2]} + r_3^{[2]} = N_2 W$. On the other hand, $M_2' W_2 + \min(M_1, N_2) W_1 = M_1 W_1 + M_2' W_2 = N_2 W$ and hence the rank condition is met for receiver two. This completes the proof for $W_1 \geq W_2$.

- $W_1 < W_2$

Substituting $i_{\max}=2$ and $i_{\min}=1$ in (9), we have

$$\begin{aligned}
r_1^{[1]} &= \min\{N_1, M_1+M_2'\}\, W_1 \overset{(a)}{=} N_1W_1\\
r_2^{[1]} &= \min\Big\{N_1(W_2-W_1), M_2'(W_2-W_1)+\min\Big(\ M_1(W_2-W_1), M_1W_1, N_2W_1\Big)\Big\} \overset{(b)}{=} N_1(W_2-W_1)\\
r_3^{[1]} &= \min\Big\{N_1(W-W_2), \min\Big(\ M_1(W-W_2), M_1W_1, N_2W_1\Big)+\min\Big(\ M_2'(W-W_2), M_2'W_2, N_1W_2\Big)\Big\}\\
&\overset{(c)}{=} \min\Big\{N_1(W-W_2), M_1W_1+\min\Big(\ M_2'(W-W_2), N_1W_2\Big)\Big\} \overset{(d)}{=} N_1(W-W_2),
\end{aligned}$$

where (a), (b), and (c) follow from the assumptions $N_1 < M_1 \leq N_2 < M_2' \leq N_1+N_2$ and $W-W_2 > W_1$ and (d) follows from $N_1 < M_2'$ and the fact that $M_1W_1+N_1W_2 = N_1W$. Therefore, $r_1^{[1]}+r_2^{[1]}+r_3^{[1]} = N_1W$. On the other hand, $M_1W_1+\min(M_2', N_1)W_2 = M_1W_1+N_1W_2 = N_1W$ and hence the rank condition is met for receiver one. For receiver two, we have

$$\begin{aligned}
r_1^{[2]} &= \min\{N_2, M_1+M_2'\}\, W_1 \overset{(a)}{=} N_2W_1\\
r_2^{[2]} &= \min\Big\{N_2(W_2-W_1), M_2'(W_2-W_1)+\min\Big(\ M_1(W_2-W_1), M_1W_1, N_2W_1\Big)\Big\} \overset{(b)}{=} N_2(W_2-W_1)\\
r_3^{[2]} &= \min\Big\{N_2(W-W_2), \min\Big(\ M_2'(W-W_2), M_2'W_2, N_1W_2\Big)+\min\Big(\ M_1(W-W_2), M_1W_1, N_2W_1\Big)\Big\}\\
&\overset{(c)}{=} \min\Big\{N_2(W-W_2), \min\Big(\ M_2'(W-W_2), N_1W_2\Big)+M_1W_1\Big\} \overset{(d)}{=} N_2(W-W_2),
\end{aligned}$$

where (a), (b), and (c) follow from the assumptions $N_1 < M_1 \leq N_2 < M_2' \leq N_1+N_2$ and $W-W_2 > W_1$ and (d) follows from $N_2 < M_2'$ and the fact that $M_1W_1+N_1W_2 = N_1W > N_2(W-W_2)$ where the reason for the last inequality simply follows from $\frac{W}{W_2} = \frac{M_2'-N_1}{N_2-N_1} \leq \frac{N_2}{N_2-N_1}$. Therefore, $r_1^{[2]}+r_2^{[2]}+r_3^{[2]} = N_2W$. On the other hand, $M_2'W_2+\min(M_1, N_2)W_1 = M_2'W_2+M_1W_1 = N_2W$ and hence the rank condition is met for receiver two. This completes the proof.

## APPENDIX C
## PROOF OF ACHIEVABILITY FOR CLASS $\mathcal{C}_4$

To show point $P_3 = (\frac{M_1'N_1(M_2'-N_2)}{M_1'M_2'-N_1N_2}, \frac{M_2'N_2(M_1'-N_1)}{M_1'M_2'-N_1N_2})$ is achievable, we show that $W = M_1'M_2'-N_1N_2$, $W_1 = N_1(M_2'-N_2)$, and $W_2 = N_2(M_1'-N_1)$ satisfy the rank conditions in (9). We have

$$\begin{aligned}
r_1^{[1]} &= \min\{N_1, M_1'+M_2'\}\, W_{i_{\min}} \overset{(a)}{=} N_1W_{i_{\min}}\\
r_2^{[1]} &= \min\Big\{N_1(W_{i_{\max}}-W_{i_{\min}}), M'_{i_{\max}}(W_{i_{\max}}-W_{i_{\min}})+\min\Big(\ M'_{i_{\min}}(W_{i_{\max}}-W_{i_{\min}}), M'_{i_{\min}}W_{i_{\min}}, N_{i_{\max}}W_{i_{\min}}\Big)\Big\}\\
&\overset{(b)}{=} N_1(W_{i_{\max}}-W_{i_{\min}})
\end{aligned}$$

$$
\begin{aligned}
r_3^{[1]} &= \min\Big\{N_1(W - W_{i_{\max}}), \min\Big(\, M_1'(W - W_{i_{\max}}), M_1'W_1, N_2W_1\Big) + \min\Big(\, M_2'(W - W_{i_{\max}}), M_2'W_2, N_1W_2\Big)\Big\} \\
&\overset{(b)}{=} \min\Big\{N_1(W - W_{i_{\max}}), \min\Big(\, M_1'(W - W_{i_{\max}}), N_2W_1\Big) + \min\Big(\, M_2'(W - W_{i_{\max}}), N_1W_2\Big)\Big\} \\
&\overset{(c)}{=} N_1(W - W_{i_{\max}}) \\
r_1^{[2]} &= \min\{N_2, M_1' + M_2'\}\, W_{i_{\min}} \overset{(a)}{=} N_2W_{i_{\min}}
\end{aligned}
$$

$$
\begin{aligned}
r_2^{[2]} &= \min\Big\{N_2(W_{i_{\max}} - W_{i_{\min}}), M_{i_{\max}}'(W_{i_{\max}} - W_{i_{\min}}) + \min\Big(\, M_{i_{\min}}'(W_{i_{\max}} - W_{i_{\min}}), M_{i_{\min}}'W_{i_{\min}}, N_{i_{\max}}W_{i_{\min}}\Big)\Big\} \\
&\overset{(b)}{=} N_2(W_{i_{\max}} - W_{i_{\min}}) \\
r_3^{[2]} &= \min\Big\{N_2(W - W_{i_{\max}}), \min\Big(\, M_2'(W - W_{i_{\max}}), M_2'W_2, N_1W_2\Big) + \min\Big(\, M_1'(W - W_{i_{\max}}), M_1'W_1, N_2W_1\Big)\Big\} \\
&\overset{(b)}{=} \min\Big\{N_2(W - W_{i_{\max}}), \min\Big(\, M_2'(W - W_{i_{\max}}), N_1W_2\Big) + \min\Big(\, M_1'(W - W_{i_{\max}}), N_2W_1\Big)\Big\} \\
&\overset{(c)}{=} N_2(W - W_{i_{\max}}),
\end{aligned}
$$

where (a) and (b) follow from $\min(M_1', M_2') > N_2 > N_1$. To prove (c), we need to prove that $N_1W_2 + N_2W_1 > N_2(W - W_{i_{\max}})$. In fact, we can prove the following inequality which is stronger than what is required

$$
N_1W_2 + N_2W_1 \geq N_2(W - W_2). \tag{23}
$$

To this aim, we notice that $W = W_1 + \frac{M_2'}{N_2}W_2$ and therefore (23) is reduced to $M_2' \leq N_1 + N_2$ which is obvious. Therefore, we have $r_1^{[i]} + r_2^{[i]} + r_3^{[i]} = N_iW$, $i = 1, 2$. On the other hand, one can easily check that

$$
\begin{aligned}
M_1'W_1 + \min(M_2', N_1)W_2 &= M_1'W_1 + N_1W_2 = N_1W \\
M_2'W_2 + \min(M_1', N_2)W_1 &= M_2'W_2 + N_2W_1 = N_2W,
\end{aligned}
$$

and therefore the rank condition is met at both receivers. This completes the proof.

## APPENDIX D
## PROOF OF ACHIEVABILITY FOR CLASS $\mathcal{C}_5$

To show point $S_3 = (M_1, \frac{M_2(N_1 - M_1)}{N_1})$ is achievable, we show that $W = W_1 = N_1$ and $W_2 = N_1 - M_1$ satisfy the rank conditions in (9). Since $W_1 > W_2$, we have $i_{\max} = 1$ and $i_{\min} = 2$. Substituting in (9), we have

$$
\begin{aligned}
r_1^{[1]} &= \min\{N_1, M_1 + M_2\}\, W_2 = N_1(N_1 - M_1) \\
r_2^{[1]} &= \min\Big\{N_1M_1, M_1^2 + \min\Big(\, M_2M_1, M_2(N_1 - M_1), N_1(N_1 - M_1)\Big)\Big\} = N_1M_1 \\
r_3^{[1]} &= 0,
\end{aligned}
$$

and hence $r_1^{[1]}+r_2^{[1]}+r_3^{[1]} = N_1^2$. On the other hand $M_1W_1+\min(M_2,N_1)W_2 = M_1N_1+N_1(N_1-M_1) = N_1^2$ and therefore the rank condition is verified for receiver one. For receiver two, we have

$$M_2W_2 + \min(M_1, N_2)W_1 = M_2(N_1 - M_1) + M_1N_1,$$

and

$$\begin{aligned}
r_1^{[2]} &= \min\{N_2, M_1 + M_2\} W_2 = \min\{N_2, M_1 + M_2\}(N_1 - M_1)\\
r_2^{[2]} &= \min\Big\{N_2M_1, M_1^2 + \min\Big( M_2M_1, M_2(N_1 - M_1), N_1(N_1 - M_1)\Big)\Big\}\\
&= M_1 \min\Big\{N_2, M_1 + \min\Big(M_2, \tfrac{N_1(N_1-M_1)}{M_1}\Big)\Big\}\\
r_3^{[2]} &= 0.
\end{aligned}$$

It is easy to check that $M_1 + \min(M_2, \frac{N_1(N_1-M_1)}{M_1}) > N_1$, and therefore $r_2^{[2]} > M_1N_1$. Hence

$$r_1^{[2]} + r_2^{[2]} + r_3^{[2]} > \min\{N_2, M_1 + M_2\}(N_1 - M_1) + M_1N_1 > M_2(N_1 - M_1) + M_1N_1.$$

Therefore, the rank condition is satisfied for receiver two and the proof is complete.

## Appendix E
## Proof of achievability for Subclass $\mathcal{C}_{61}$ and Subclass $\mathcal{C}_{62}$

In this part, we show that point $T_4 = (M_1, N_2 - M_1)$ is achievable for subclass $\mathcal{C}_{61}$ and subclass $\mathcal{C}_{62}$:

Proof of Achievability of point $T_4$ for subclass $\mathcal{C}_{61}$

Since for this subclass $M_2 > N_1+N_2-M_1$, we can assume that transmitter two employs only $N_1+N_2-M_1$ out of its $M_2$ transmit antennas. Therefore, in the sequel, we set $M_2' = N_1 + N_2 - M_1$. To show that point $T_4$ is achievable, we show that $W = W_1 = N_1 + N_2 - M_1$ and $W_2 = N_2 - M_1$ satisfy the rank conditions in (9). Since $W_1 > W_2$, we have $i_{\text{max}} = 1$ and $i_{\text{min}} = 2$. Substituting in (9), we have

$$\begin{aligned}
r_1^{[1]} &= \min\{N_1, M_1 + M_2'\} W_2 = N_1(N_2 - M_1)\\
r_2^{[1]} &= \min\Big\{N_1^2, M_1N_1 + \min\Big( M_2'N_1, M_2'(N_2 - M_1), N_1(N_2 - M_1)\Big)\Big\} = N_1^2\\
r_3^{[1]} &= 0,
\end{aligned}$$

and hence $r_1^{[1]} + r_2^{[1]} + r_3^{[1]} = N_1^2 + N_1(N_2 - M_1)$. On the other hand $M_1W_1 + \min(M_2', N_1)W_2 = M_1(N_1 + N_2 - M_1) + N_1(N_2 - M_1)$. Since $M_1 \leq \Delta$, we have $N_1^2 > M_1(N_1 + N_2 - M_1)$ and hence the rank condition is verified for receiver one. For receiver two, we have

$$\begin{aligned} r_1^{[2]} &= \min\{N_2, M_1 + M_2'\} W_2 = N_2(N_2 - M_1) \\ r_2^{[2]} &= \min\Big\{N_1N_2, M_1N_1 + \min\Big( M_2'N_1, M_2'(N_2 - M_1), N_1(N_2 - M_1)\Big)\Big\} = N_1N_2 \\ r_3^{[2]} &= 0. \end{aligned}$$

and hence $r_1^{[2]} + r_2^{[2]} + r_3^{[2]} = N_2(N_1 + N_2 - M_1)$. On the other hand $M_2W_2 + \min(M_1, N_2)W_1 = (N_1 + N_2 - M_1)(N_2 - M_1) + M_1(N_1 + N_2 - M_1) = N_2(N_1 + N_2 - M_1)$. Therefore, the rank condition is satisfied for receiver two and the proof is complete.

<u>Proof of Achievability of point $T_4$ for subclass $\mathcal{C}_{62}$</u>

We show that $W = W_1 = M_2$ and $W_2 = N_2 - M_1$ satisfy the rank conditions in (9). Since $W_1 > W_2$, we have $i_{\text{max}} = 1$ and $i_{\text{min}} = 2$. Substituting in (9), we have

$$\begin{aligned} r_1^{[1]} &= \min\{N_1, M_1 + M_2\} W_2 = N_1(N_2 - M_1) \\ r_2^{[1]} &= \min\Big\{ N_1(M_1 + M_2 - N_2), M_1(M_1 + M_2 - N_2) + \\ &\quad \min\Big( M_2(M_1 + M_2 - N_2), M_2(N_2 - M_1), N_1(N_2 - M_1)\Big)\Big\} = N_1(M_1 + M_2 - N_2) \\ r_3^{[1]} &= 0, \end{aligned}$$

and hence $r_1^{[1]} + r_2^{[1]} + r_3^{[1]} = N_1M_2$. On the other hand $M_1W_1 + \min(M_2, N_1)W_2 = M_1M_2 + N_1(N_2 - M_1)$. Since $M_1 \leq \Delta' = \frac{N_1(M_2-N_2)}{M_2-N_1}$, we have $M_1M_2 < N_1(M_1 + M_2 - N_2)$ and hence the rank condition is verified for receiver one. For receiver two, we have

$$\begin{aligned} r_1^{[2]} &= \min\{N_2, M_1 + M_2\} W_2 = N_2(N_2 - M_1) \\ r_2^{[2]} &= \min\Big\{ N_2(M_1 + M_2 - N_2), M_1(M_1 + M_2 - N_2) + \\ &\quad \min\Big( M_2(M_1 + M_2 - N_2), M_2(N_2 - M_1), N_1(N_2 - M_1)\Big)\Big\} = N_2(M_1 + M_2 - N_2) \\ r_3^{[2]} &= 0. \end{aligned}$$

and hence $r_1^{[2]} + r_2^{[2]} + r_3^{[2]} = N_2M_2$. On the other hand $M_2W_2 + \min(M_1, N_2)W_1 = M_2(N_2 - M_1) + M_1M_2 = N_2M_2$. Therefore, the rank condition is satisfied for receiver two and the proof is complete.

## Appendix F

## Proof of achievability for Subclass $\mathcal{C}_{63}$

In this part, we show that point $T_5 = (\frac{N_1(M_2-N_2)}{M_2-N_1}, \frac{M_2(N_2-N_1)}{M_2-N_1})$ and $T_6 = (M_1, \frac{M_2(N_1-M_1)}{N_1})$ are achievable for subclass $\mathcal{C}_{63}$. Remember that for for subclass $\mathcal{C}_{63}$ we have $\Delta' \leq M_1 < N_1 < N_2 < M_2 < N_1 + N_2 - M_1$.

**<u>Proof of Achievability for point $T_5$</u>**

We will show that $W = M_1(M_2 - N_1)$, $W_1 = N_1(M_2 - N_2)$, and $W_2 = M_1(N_2 - N_1)$ satisfy the rank conditions in (9). First, we notice that $W = \frac{M_1}{N_1}W_1 + W_2$ and therefore $W < W_1 + W_2$. To this end, we differentiate between two cases:

- $W_1 \geq W_2$

For this case, we have $i_{\text{max}} = 1$ and $i_{\text{min}} = 2$. Substituting in (9), we have

$$
\begin{aligned}
r_1^{[1]} &= \min\{N_1, M_1 + M_2\}\, W_2 \stackrel{(a)}{=} N_1 W_2 \\
r_2^{[1]} &= \min\Big\{N_1(W_1 - W_2), M_1(W_1 - W_2) + \min\Big(\, M_2(W_1 - W_2), M_2W_2, N_1W_2\Big)\Big\} \\
&\stackrel{(b)}{=} \min\Big\{N_1(W_1 - W_2), M_1(W_1 - W_2) + \min\Big(\, M_2(W_1 - W_2), N_1W_2\Big)\Big\} \stackrel{(c)}{=} N_1(W_1 - W_2) \\
r_3^{[1]} &= \min\Big\{N_1(W - W_1), \min\Big(\, M_1(W - W_1), M_1W_1, N_2W_1\Big) + \min\Big(\, M_2(W - W_1), M_2W_2, N_1W_2\Big)\Big\} \\
&\stackrel{(d)}{=} \min\Big\{N_1(W - W_1), M_1(W - W_1) + \min\Big(\, M_2(W - W_1), N_1W_2\Big)\Big\} \stackrel{(e)}{=} N_1(W - W_1),
\end{aligned}
$$

where (a), (b), and (d) follow from the assumptions $M_1 < N_1 < N_2 < M_2$ and $W - W_1 < W_2 \leq W_1$, and (e) follows from the obvious inequality $N_1W_2 > (N_1 - M_1)(W - W_1)$. To prove (c), we need to prove that $M_1(W_1 - W_2) + N_1W_2 \geq N_1(W_1 - W_2)$. Equivalently, we need to prove that $\frac{W_1}{W_2} - 1 \leq \frac{N_1}{N_1 - M_1}$. Since $M_1 > \Delta' = \frac{N_1(M_2 - N_2)}{M_2 - N_1}$, it follows that

$$
\begin{aligned}
\frac{M_1(N_2 - N_1)}{N_1(M_2 - N_2)} > \frac{N_2 - N_1}{M_2 - N_1} &\Rightarrow \frac{W_2}{W_1} > \frac{N_2 - N_1}{M_2 - N_1} \\
\Rightarrow \frac{W_1}{W_2} - 1 < \frac{M_2 - N_1}{N_2 - N_1} - 1 = \frac{M_2 - N_2}{N_2 - N_2} &= \frac{M_1}{N_1}\frac{W_1}{W_2} \\
\Rightarrow (1 - \frac{M_1}{N_1})\frac{W_1}{W_2} < 1 \Rightarrow \frac{W_1}{W_2} &< \frac{N_1}{N_1 - M_1},
\end{aligned}
$$

which is the desired result. Therefore, $r_1^{[1]} + r_2^{[1]} + r_3^{[1]} = N_1W$. On the other hand, $M_1W_1 + \min(M_2', N_1)W_2 = M_1W_1 + N_1W_2 = N_1W$ and hence the rank condition is met for receiver one. For receiver two, we have

$$
\begin{aligned}
r_1^{[2]} &= \min\{N_2, M_1 + M_2\}\, W_2 \stackrel{(a)}{=} N_2 W_2 \\
r_2^{[2]} &= \min\Big\{N_2(W_1 - W_2), M_1(W_1 - W_2) + \min\Big(\, M_2(W_1 - W_2), M_2W_2, N_1W_2\Big)\Big\} \\
&\stackrel{(b)}{=} \min\Big\{N_2(W_1 - W_2), M_1(W_1 - W_2) + \min\Big(\, M_2(W_1 - W_2), N_1W_2\Big)\Big\} \stackrel{(c)}{=} N_2(W_1 - W_2) \\
r_3^{[2]} &= \min\Big\{N_2(W - W_1), \min\Big(\, M_2(W - W_1), M_2W_2, N_1W_2\Big) + \min\Big(\, M_1(W - W_1), M_1W_1, N_2W_1\Big)\Big\} \\
&\stackrel{(d)}{=} \min\Big\{N_2(W - W_1), \min\Big(\, M_2(W - W_1), N_1W_2\Big) + M_1(W - W_1)\Big\} \stackrel{(e)}{=} N_2(W - W_1),
\end{aligned}
$$

where (a), (b), and (d) follow from the assumption $M_1 < N_1 < N_2 < M_2$ and $W - W_1 < W_2 \leq W_1$. To prove (c), we need to show that $M_1(W_1 - W_2) + N_1W_2 \geq N_2(W_1 - W_2)$ or equivalently $\frac{W_1}{W_2} - 1 < \frac{N_1}{N_2 - M_1}$. To prove (e), we need to show that $N_1W_2 + M_1(W - W_1) > N_2(W - W_1)$ or equivalently $\frac{W - W_1}{W_2} < \frac{N_1}{N_2 - M_1}$.

First, we prove that $\frac{W_1}{W_2} - 1 < \frac{N_1}{N_2 - M_1}$. The right hand side of this inequality can be written as

$$\frac{W_1}{W_2} - 1 = \frac{N_1(M_2 - N_2)}{M_1(N_2 - N_1)} - 1 < \frac{N_1(M_2 - N_2)}{\Delta'(N_2 - N_1)} - 1, \tag{24}$$

where the last inequality follows from $M_1 > \Delta'$. After some algebraic manipulation, one can prove that $\frac{N_1(M_2-N_2)}{\Delta'(N_2-N_1)} - 1 = \frac{\Delta'}{N_1 - \Delta'}$. On the other hand, from $M_2 < N_1 + N_2 - M_1$, it follows that $\Delta' < \Delta = \frac{N_1(N_1-M_1)}{N_2-M_1}$. Since $M_1 > \Delta'$ and $\Delta$ is a decreasing function in terms of $M_1$, we will have

$$\Delta' < \frac{N_1(N_1 - M_1)}{N_2 - M_1} < \frac{N_1(N_1 - \Delta')}{N_2 - \Delta'} \;\Rightarrow\; \frac{\Delta'}{N_1 - \Delta'} < \frac{N_1}{N_2 - \Delta'}. \tag{25}$$

By combining (24) and (25) and noting that $M_1 > \Delta'$, we reach to

$$\frac{W_1}{W_2} - 1 < \frac{N_1}{N_2 - \Delta'} < \frac{N_1}{N_2 - M_1},$$

which is the desired result. We then prove that $\frac{W-W_1}{W_2} < \frac{N_1}{N_2-M_1}$. Since $W = \frac{M_1}{N_1}W_1 + W_2$ and $W_1 \geq W_2$, we have

$$W - W_1 = (\frac{M_1}{N_1} - 1)W_1 + W_2 \leq (\frac{M_1}{N_1} - 1)W_2 + W_2 = \frac{M_1}{N_1}W_2 \Rightarrow \frac{W - W_1}{W_2} \leq \frac{M_1}{N_1}. \tag{26}$$

On the other hand, we have

$$\frac{M_1}{N_1} \overset{(a)}{\leq} \frac{M_2 - N_2}{N_2 - N_1} \overset{(b)}{\leq} \frac{N_1 - M_1}{N_2 - N_1}, \tag{27}$$

where (a) follows from $W_1 \geq W_2$ and (b) follows from $M_2 < N_1 + N_2 - M_1$. After some manipulation, (27) is reduced to $\frac{M_1}{N_1} \leq \frac{N_1}{N_2}$. Therefore, from (26), we have

$$\frac{W - W_1}{W_2} \leq \frac{N_1}{N_2} < \frac{N_1}{N_2 - M_1},$$

which is the desired result. Therefore, $r_1^{[2]} + r_2^{[2]} + r_3^{[2]} = N_2 W$. On the other hand, $M_2 W_2 + \min(M_1, N_2) W_1 = M_1 W_1 + M_2 W_2 = N_2 W$ and hence the rank condition is met for receiver two. This completes the proof for $W_1 \geq W_2$.

- $W_1 < W_2$

For this case, we have $i_{\text{max}} = 2$ and $i_{\text{min}} = 1$. Substituting in (9), we have

$$\begin{aligned}
r_1^{[1]} &= \min\{N_1, M_1 + M_2\} W_1 \overset{(a)}{=} N_1 W_1 \\
r_2^{[1]} &= \min\Big\{N_1(W_2 - W_1), M_2(W_2 - W_1) + \min\Big( M_1(W_2 - W_1), M_1 W_1, N_2 W_1\Big)\Big\} \overset{(b)}{=} N_1(W_2 - W_1) \\
r_3^{[1]} &= \min\Big\{N_1(W - W_2), \min\Big( M_1(W - W_2), M_1 W_1, N_2 W_1\Big) + \min\Big( M_2(W - W_2), M_2 W_2, N_1 W_2\Big)\Big\} \\
&\overset{(c)}{=} \min\Big\{N_1(W - W_2), M_1(W - W_2) + \min\Big( M_2(W - W_2), N_1 W_2\Big)\Big\} \overset{(d)}{=} N_1(W - W_2),
\end{aligned}$$

where (a), (b), and (c) follow from the assumptions $N_1 < M_1 \leq N_2 < M_2' \leq N_1 + N_2$ and $W - W_2 < W_1$ and (d) follows from $N_1 < M_2$ and the the obvious inequality $N_1 W_2 > (N_1 - M_1)(W - W_2)$. Therefore, $r_1^{[1]} + r_2^{[1]} + r_3^{[1]} = N_1 W$. On the other hand, $M_1 W_1 + \min(M_2, N_1) W_2 = M_1 W_1 + N_1 W_2 = N_1 W$ and hence the rank condition is met for receiver one. For receiver two, we have

$$\begin{aligned}
r_1^{[2]} &= \min\{N_2, M_1 + M_2\} W_1 \overset{(a)}{=} N_2 W_1 \\
r_2^{[2]} &= \min\Big\{N_2(W_2 - W_1), M_2(W_2 - W_1) + \min\Big(\ M_1(W_2 - W_1), M_1 W_1, N_2 W_1\Big)\Big\} \overset{(b)}{=} N_2(W_2 - W_1) \\
r_3^{[2]} &= \min\Big\{N_2(W - W_2), \min\Big(\ M_2(W - W_2), M_2 W_2, N_1 W_2\Big) + \min\Big(\ M_1(W - W_2), M_1 W_1, N_2 W_1\Big)\Big\} \\
&\overset{(c)}{=} \min\Big\{N_2(W - W_2), \min\Big(\ M_2(W - W_2), N_1 W_2\Big) + M_1(W - W_2)\Big\} \overset{(d)}{=} N_2(W - W_2),
\end{aligned}$$

where (a), (b), and (c) follow from the assumptions $N_1 < M_1 \leq N_2 < M_2' \leq N_1 + N_2$ and $W - W_2 < W_1$. To prove (d), we need to show that $N_1 W_2 + M_1(W - W_2) > N_2(W - W_2)$ or equivalently

$$\frac{W - W_2}{W_2} < \frac{N_1}{N_2 - M_1}. \tag{28}$$

The right hand side of (28) is equal to $\frac{M_1 W_1}{N_1 W_2}$. From $W_2 > W_1$, it follows that $M_1(N_2 - N_1) > N_1(M_2 - N_2)$ or equivalently $M_1 > M_1^*$ where $M_1^* \triangleq \frac{N_1(M_2 - N_2)}{N_2 - N_1}$. Therefore, (28) can be equivalently expresses as

$$M_1^* < \frac{N_1^2}{N_2 - M_1}. \tag{29}$$

From $M_2 < N_1 + N_2 - M_1$ and $M_1 > M_1^*$, it follows that $M_2 < N_1 + N_2 - M_1^*$ and therefore $M_2 - N_2 < N_1 - M_1^*$. It then follows that $M_1^* < \frac{N_1(N_1 - M_1^*)}{N_2 - N_1}$ which is equivalent to $M_1^* < \frac{N_1^2}{N_2}$. Therefore $M_1^* < \frac{N_1^2}{N_2 - M_1}$ and (29) is valid. Therefore, $r_1^{[2]} + r_2^{[2]} + r_3^{[2]} = N_2 W$. On the other hand, $M_2 W_2 + \min(M_1, N_2) W_1 = M_2 W_2 + M_1 W_1 = N_2 W$ and hence the rank condition is met for receiver two. This completes the proof.

**Proof of Achievability for point $T_6$**

To show point $T_6 = (M_1, \frac{M_2(N_1 - M_1)}{N_1})$ is achievable, we show that $W = W_1 = N_1$ and $W_2 = N_1 - M_1$ satisfy the rank conditions in (9). Since $W_1 > W_2$, we have $i_{\max} = 1$ and $i_{\min} = 2$. Substituting in (9), we have

$$\begin{aligned}
r_1^{[1]} &= \min\{N_1, M_1 + M_2\} W_2 \overset{(a)}{=} N_1 W_2 \\
r_2^{[1]} &= \min\Big\{N_1(W_1 - W_2), M_1(W_1 - W_2) + \min\Big(\ M_2(W_1 - W_2), M_2 W_2, N_1 W_2\Big)\Big\} \overset{(b)}{=} N_1(W_1 - W_2) \\
r_3^{[1]} &= 0,
\end{aligned}$$

where (a) follows from $N_1 < M_2$ and (b) follows from $M_1 < N_1 < M_2$ and the fact that $M_1(W_1 - W_2) + N_1 W_2 > N_1(W_1 - W_2)$. Hence $r_1^{[1]} + r_2^{[1]} + r_3^{[1]} = N_1 W_1 = N_1^2$. On the other hand $M_1 W_1 + \min(M_2, N_1) W_2 = M_1 N_1 + N_1(N_1 - M_1) = N_1^2$ and therefore the rank condition is verified for receiver one. For receiver two, we have

$$M_2W_2 + \min(M_1, N_2)W_1 = M_2(N_1 - M_1) + M_1N_1,$$

and

$$\begin{aligned} r_1^{[2]} &= \min\{N_2, M_1 + M_2\}W_2 \overset{(a)}{=} N_2(N_1 - M_1) \\ r_2^{[2]} &= \min\Big\{N_2(W_1 - W_2), M_1(W_1 - W_2) + \min\Big( M_2(W_1 - W_2), M_2W_2, N_1W_2\Big)\Big\} \\ &\overset{(b)}{=} M_1 \min\Big\{N_2, M_1 + \min\Big(M_2, \tfrac{N_1(N_1-M_1)}{M_1}\Big)\Big\} \overset{(c)}{=} M_1 \min\Big\{N_2, M_1 + \tfrac{N_1(N_1-M_1)}{M_1}\Big\} \\ r_3^{[2]} &= 0, \end{aligned}$$

where (a), (b), and (c) follow from $N_1 < N_2 < M_2$. Now, we need to differentiate between two cases:

- $N_2 \geq M_1 + \frac{N_1(N_1-M_1)}{M_1}$
  In this case, $r_2^{[2]} = M_1^2 + N_1(N_1 - M_1)$ and therefore $r_1^{[2]} + r_2^{[2]} = M_1^2 + N_1(N_1 - M_1) + N_2(N_1 - M_1)$. On the other hand, from $M_2 < N_1 + N_2 - M_1$, it follows that

$$M_2W_2 + \min(M_1, N_2)W_1 < (N_1 + N_2 - M_1)W_2 + M_1W_1 = r_1^{[2]} + r_2^{[2]},$$

  and therefore the proof is complete for this case.

- $N_2 < M_1 + \frac{N_1(N_1-M_1)}{M_1}$
  In this case, $r_2^{[2]} = M_1N_2$ and therefore $r_1^{[2]} + r_2^{[2]} = N_1N_2$. On the other hand, from $M_1 > \Delta'$, it follows that

$$M_1(M_2 - N_1) > N_1(M_2 - N_2) \Rightarrow N_1N_2 > M_2(N_1 - M_1) + M_1N_1,$$

  which is the desired result. This completes the proof.